\documentclass[aps,showpacs,amsmath,amssymb,preprintnumbers,11pt]{revtex4-1}
\usepackage{hyperref}
\usepackage{comment}
\usepackage{xcolor}
\usepackage{mathrsfs}

\def\be{\begin{equation}}
\def\ee{\end{equation}}
\def\bea{\begin{eqnarray}}
\def\eea{\end{eqnarray}}

\newcommand{\eps}{\epsilon}

\begin{document}

\preprint{QMUL-PH-26-27}

\title{Mass/electric versus NUT/magnetic charges: \\ duality from scattering amplitudes in $D\geq4$ and for all bosonic spins}
\author{Ricardo Monteiro}
\email{ricardo.monteiro@qmul.ac.uk}
\author{Lecheng Ren}
\email{lecheng.ren@qmul.ac.uk}
\author{Daniel Siretanu}
\email{d.siretanu@qmul.ac.uk}
\affiliation{Centre for Theoretical Physics, Department of Physics and Astronomy, \\
Queen Mary University of London, Mile End Road, London E1 4NS, United Kingdom}

\begin{abstract}
We revisit Kerr-NUT metrics and related spin-$s$ fields in $D\geq4$, and study their associated scattering amplitudes. Starting in position space, we highlight the interpretation of mass and (multiple) NUT charges as being associated to distinct solutions of the rotation-deformed radial equation, which is made explicit in Cartesian multi-Kerr-Schild coordinates. This interpretation extends to electromagnetism with magnetic-type charges, and also extends to higher-spin counterparts, in accordance with the classical double or multi copy. We then establish a notion of ``electric-magnetic" duality in higher dimensions, relating mass/electric to NUT/magnetic charges, which generalises the $D=4$ case in a novel manner. For $D\geq6$, this involves the choice where the multiple magnetic charges are equal. The duality is revealed in momentum space, by the 3-point scattering amplitudes that generate the solutions. For all spins, these amplitudes are constructed from a spin-raising operator $\mathcal S_\mu$ and take the form $\varepsilon^{\mu_1\cdots\mu_s}{\mathcal S}_{\mu_1}\cdots {\mathcal S}_{\mu_s}$ acting on a scalar seed. The scalar seed of the electric sector is dual to that of the magnetic sector: where the former's rotation dependence resums to a Bessel function $J_{\frac{D-5}{2}}$, the latter resums to a Bessel function $J_{-\frac{D-5}{2}}$. Finally, we explore the notion of self-duality that arises from this picture. Studying the classical $2\!\mapsto\!2$ amplitudes that determine leading-order scattering, we find no evidence of a higher-dimensional analogue of the integrability of $D=4$ self-dual gravity.
\end{abstract} 

\keywords{Scattering amplitudes, black holes, higher-dimensional gravity}

\maketitle

\tableofcontents

\section{Introduction and summary}
\label{sec:intro}

The Kerr(-Taub)-NUT family of black holes in four dimensions \cite{Kerr:1963ud,Taub:1950ez,Newman:1963yy,1966BAPSS..14..653D} is not only of interest to phenomenology, but has also been a laboratory for ideas of duality and integrability. A recent approach to these metrics is to consider the associated scattering amplitudes. The 3-point amplitude with one graviton leg is essentially the Fourier transform of the stationary solution (more precisely, of its deviation from flatness), while the classical $2\!\mapsto\!2$ amplitude with one graviton exchange (which effectively factorises into 3-point amplitudes) encodes the effects of black hole scattering at leading perturbative order. See e.g.~\cite{Bern:2022wqg,Kosower:2022yvp,Bjerrum-Bohr:2022blt,Buonanno:2022pgc,DiVecchia:2023frv} for reviews of this approach to gravity, and in particular \cite{Guevara:2018wpp,Chung:2018kqs,Arkani-Hamed:2019ymq,Huang:2019cja,Moynihan:2020gxj,Csaki:2020inw,Emond:2020lwi,Kim:2020cvf,Monteiro:2020plf,Crawley:2021auj,Monteiro:2021ztt,Doran:2026bng,Caron-Huot:2018ape,Bianchi:2024shc,Bianchi:2025xol} for work on the 3-point amplitudes that our paper builds on. While the amplitudes-based approach is typically perturbative, the multi-Kerr-Schild nature of the Kerr-NUT family facilitates the map between a 3-point amplitude and the corresponding exact solution.

Duality and integrability features are elegantly revealed in $D=4$ in terms of the scattering amplitudes, so it is natural to explore the amplitudes for insights into $D>4$. Among the theories of interest, we have not just gravity but also the theories related to it by the classical double copy \cite{Monteiro:2014cda,Luna:2015paa} and its higher-spin extensions \cite{Didenko:2022qxq}. Our starting point is the recent construction in \cite{Bianchi:2024shc,Bianchi:2025xol} of 3-point amplitudes for the $D\geq4$ extension of the Kerr solution representing a rotating black hole; this extension is known as the Myers-Perry family \cite{Myers:1986un}. We will study here the $D\geq4$ extension of the Kerr-NUT solution, introduced by Chen, L\"u and Pope \cite{Chen:2006xh}. This family shares remarkable properties of the 4D case: algebraic type D \cite{Hamamoto:2006zf,Mason:2010zzc} in a higher-dimensional extension  \cite{Coley:2004jv,Ortaggio:2012jd} of the Petrov classification; multi-Kerr-Schild type \cite{Chen:2007fs}; complete integrability of geodesic motion \cite{Page:2006ka}; and separability of the Hamilton-Jacobi, Klein-Gordon, and Dirac equations \cite{Frolov:2006pe}. These properties are related to the existence of a `principal tensor', as reviewed in \cite{Frolov:2017kze}. The family admits as limiting cases a variety of higher-dimensional Taub-NUT metrics \cite{Mann:2003zh,Mann:2005ra,Clarkson:2006zk,Chen:2006ea,Krtous:2015zco}. In fact, it also includes a cosmological constant $\Lambda$, generalising the zero-NUT case of \cite{Gibbons:2004js,Hawking:1998kw}. Because we will be interested in the connection to scattering amplitudes, we will set $\Lambda=0$ in the main body of the paper, but will include $\Lambda$ in an appendix for completeness. We expect the momentum-space statements to have a counterpart in (A)dS, but will not pursue this here. 

Overall, the Kerr-NUT family provides a rich playground where one finds both unexpected similarities and important distinctions from the four-dimensional case. Let us summarise our findings.
\\

{\bf Summary}
\begin{itemize}
    \item We will associate mass and NUT charges to alternative solutions for $r^2$ of the spheroidal (i.e.~rotation-deformed) radial equation, with each solution contributing one Kerr-Schild monomial to the metric. This is closely related to previous coordinate systems for the Kerr-NUT family, including in $D=4$, but the explicit association to the spheroidal radial equation is, to our knowledge, new. Our manifestly Minkowskian base metric will also allow us to explore more easily the connection to scattering amplitudes.
    \item We present a straightforward extension of the `electric-magnetic' duality within the Kerr-NUT family in even $D>4$, different from previously considered extensions that involve distinct types of tensor fields. In particular, we find that the pure-mass solution (Myers-Perry, or (A)dS generalisation) is dual to a pure-NUT solution whose  multiple NUT charges are equal. The duality is hidden when looking at the metric, but is revealed by the associated 3-point scattering amplitude: it acts as $J_\alpha\mapsto J_{-\alpha}$ on a seed Bessel function that encodes the multipolar structure. We note that this duality acts on the space of Kerr-NUT solutions with equal NUT charges, and does not imply a dynamical duality, even in linearised gravity.
    \item Alongside gravity, we also discuss analogous solutions in electromagnetism and in massless scalar theory, in the context of the classical double copy. The analogues of NUT charges in electromagnetism are monopole-like charges. While in position space we have the multi-Kerr-Schild double copy \cite{Luna:2015paa,Chawla:2022ogv}, in momentum space we construct a spin-raising operator $\mathcal S_\mu$ relating the 3-point amplitudes for gravity and for electromagnetism to those for the massless scalar theory, which therefore acts as the seed. Moreover, the spin-raising operator allows us to extend this story to massless bosonic spin-$s$ fields, mirroring the classical multi copy in position space \cite{Didenko:2022qxq}. It is the momentum-space structure of spin-raising operators acting on a scalar seed that makes the aforementioned duality manifest.
    \item In $D=4$, the sector of self-dual charges is part of a truncation of complexified gravity, known as self-dual gravity, which is integrable \cite{10.1093/oso/9780198534983.001.0001}. A proxy for this integrability is the vanishing of the $2\!\mapsto\!2$ tree amplitude associated with leading-order scattering \cite{Doran:2026bng}. We study the analogous problem in higher dimensions, and find no evidence that integrability holds for self-dual charges beyond $D=4$.
\end{itemize}

The paper is organised as follows. We start with position space in section~\ref{sec:KTN}. First, we discuss the Kerr-NUT family, its double- and multi-copy structure, and reality conditions. Then, we introduce the notion of duality between the mass/electric and the equal-NUT/magnetic charges, which is non-trivial in even $D$, and discuss the simplifications in the equal-rotations case. In section~\ref{sec:amps}, we move to momentum space. A review of the relation between curvatures and 3-point amplitudes is followed by the construction of scattering amplitudes from a spin-raising operator, first in the Myers-Perry sector and then in the dual NUT sector, concluding with the all-spins formulas. In section~\ref{sec:tests}, we translate the structure from momentum space back to position space, describing the spin-$s$ curvatures in terms of spin-raising operations on scalar seeds. This allows us to perform tests of our amplitude formulas, both for generic rotations and by restricting to the equal-rotations case. Moving beyond 3-point amplitudes, in section~\ref{sec:SDint} we study $2\!\mapsto\!2$ scattering to check whether the self-dual sector in higher dimensions shares the special properties of $D=4$. Finally, we conclude with a discussion of possible future directions in section~\ref{sec:conclusion}.


\section{Kerr-NUT solutions}
\label{sec:KTN}

\subsection{The Kerr-NUT metric}
\label{sec:KTNgen}

We will consider the Kerr-NUT family in $D\geq 4$ obtained in \cite{Chen:2006xh}. We will restrict here to vanishing cosmological constant, but include that generalisation, alongside various details, in appendix~\ref{app:KTNLambda}. It is useful to express
\be
D=2m+2-\epsilon, \qquad \text{where}\qquad  m = \# \text{ rotation planes},
\qquad
\epsilon =
\begin{cases}
0, & \text{even $D$,} \\
1, & \text{odd $D$.}
\end{cases}
\ee
The parameters of the family are: the mass $M$; the rotation parameters $a_i$, with $i=1,\ldots,m$; and the NUT charges $N_\alpha$, with $\alpha=1,\ldots,m-\epsilon$. The mass can be packaged together with the NUT charges as $N_0=M$. (As found in \cite{Chen:2006xh}, and as we will explain later, one of the charges $N_\alpha$ is redundant in odd $D$.) The metric admits a multi-Kerr-Schild form, originally obtained in \cite{Chen:2007fs},
\be
\label{eq:mKS}
ds^2=ds^2_\text{flat} + \sum_{\alpha=0}^{m-\epsilon}N_\alpha \,\Phi_{(\alpha)}\, \ell_{(\alpha)}^2\,.
\ee
Instead of the coordinates used in \cite{Chen:2007fs}, we will employ Cartesian coordinates for this Kerr-Schild form, such that the flat metric, the Kerr-Schild scalars $\Phi_{(\alpha)}$, and the null, mutually-orthogonal and geodesic covectors $\ell_{(\alpha)}$ are written, respectively, as
\begin{align}
\label{eq:ds2flat}
ds^2_\text{flat} & = -dt^2 + (1-\epsilon)dz^2+ \sum_{i=1}^m (dx_i^2+dy_i^2)   \,, \\
\Phi_{(\alpha)} & \equiv \Phi \big|_{r= r_{(\alpha)}}\,, \qquad   \Phi=\frac{r^{1+\epsilon}}{\prod_{i=1}^m(r^2+a_i^2)}\left( 1- \sum_{i=1}^m \frac{a_i^2(x_i^2+y_i^2)}{(r^2+a_i^2)^2}\right)^{-1}\,, 
\label{eq:KSscalar}\\
\ell_{(\alpha)} & \equiv \ell\big|_{r= r_{(\alpha)}}\,, \qquad \ell =dt + (1-\epsilon)\frac{zdz}{r}  
+ \sum_{i=1}^m \frac{r(x_idx_i+y_idy_i)+a_i(x_idy_i-y_idx_i)}{r^2+a_i^2}   \,. 
\end{align}
The `radial' variables $r_{(\alpha)}$ above are the $m+1-\epsilon$ solutions (up to a sign) to the spheroidal radial equation,
\be
\label{eq:r}
\sum_{i=1}^m \frac{x_i^2+y_i^2}{r^2+a_i^2} + (1-\epsilon)\frac{z^2}{r^2} =1\,.
\ee
As described in the appendix~\ref{app:KTNLambdasol}, the squared solutions $r_{(\alpha)}^2$ to this equation are real. If we take the rotation parameters to be generic and order them such that $a_i^2<a_{i+1}^2$, then each $r_{(\alpha)}^2$ lies in an interval as follows:
\begin{align}
\label{eq:iinteven}
& \text{for} \quad \eps=0\,, \qquad (-a_m^2,-a_{m-1}^2)\,, \; \cdots\,,\; (-a_2^2,-a_1^2)\,,\; (-a_1^2,0)\,,\; (0,+\infty)\,, \\
\label{eq:iintodd}
& \text{for} \quad \eps=1\,, \qquad (-a_m^2,-a_{m-1}^2)\,, \; \cdots\,,\; (-a_2^2,-a_1^2)\,,\; (-a_1^2,+\infty)\,,
\end{align}
matching the total of $m+1-\epsilon$ solutions. We can classify the solutions as being associated to mass and NUT charges.
\begin{itemize}
    \item We take the largest $r_{(\alpha)}^2$ to be $r_{(0)}^2$, and pick the square root such that  $r_{(0)}>0$. (To be precise, in odd $D$, $r_{(0)}^2$ is only positive when the Cartesian coordinates are outside of a region set by the $a_i^2$.) This is the usual spheroidal radius. The $\alpha=0$ Kerr-Schild monomial in the metric is the mass (Myers-Perry) monomial, with $N_0=M$.
    \item For $0<\alpha<m-\epsilon$, we have the NUT monomials. As $r_{(\alpha>0)}^2<0$, we have imaginary $r_{(\alpha>0)}$, requiring each a choice of sign. We will describe later the reality conditions on our parameters $N_\alpha$.
\end{itemize}

The conditions that the $\ell_{(\alpha)}$ are null, mutually-orthogonal and geodesic are independent of the choices of square-root signs for the $r_{(\alpha>0)}$, and so is the fact that the vacuum Einstein equations hold. This follows from the dictionary in appendix~\ref{app:KTNLambdaCL}, where we map the coordinates used in \cite{Chen:2007fs} to our Cartesian coordinates. Notice that the imaginary $r_{(\alpha)}$ will make the metric complex. Indeed, it was observed long ago that to write the $D=4$  Kerr-NUT metric in double-Kerr-Schild form, a complex coordinate transformation is required \cite{Plebanski:1976gy}; in our formulation, if $r_{(0)}$ gives the Kerr monomial, then $r_{(1)}=-iaz/r_{(0)}$ gives the NUT monomial. On the other hand, one can write a real multi-Kerr-Schild metric in $D$ dimensions by analytic continuation to signature $(\lfloor D/2\rfloor,\lfloor (D+1)/2\rfloor)=(m+1-\epsilon,m+1)$, as used in \cite{Chen:2007fs}; this corresponds to (2,2) signature in $D=4$. 
We work in Lorentzian signature, and will discuss in a dedicated section the reality conditions for the metric, which require in even $D$ that our geometric NUT parameters $N_{\alpha>0}$ are actually $i$ times the real NUT charges. 

One revealing feature is that the NUT charges are allowed as parameters of a vacuum solution to the Einstein equations for precisely the same reason as the mass is, namely that the associated $r_{(\alpha)}$ are solutions to the spheroidal radial equation. This conclusion extends to monopole-type charges in electromagnetism, as we shall see. To our knowledge, this association between NUT/magnetic charges and the spheroidal radial equation has not been explicitly discussed in the literature. It is, however, implicit in previous formulations of the Kerr-NUT metric \cite{Carter:1968ks,Chen:2006xh,Chen:2007fs}, because some coordinates there turn out to be identified with the $r_{(\alpha)}$; see the appendix~\ref{app:KTNLambdaCL} for the precise map.

Before we proceed, let us mention two features of the functions $\Phi_{(\alpha)}$ that will be important later. The first is the relation
\be
\label{eq:idPhi}
\Phi_{(\alpha)}
=
\frac{r_{(\alpha)}^{1-\epsilon}}
{\displaystyle\prod_{\beta\neq\alpha}
(r_{(\alpha)}^2-r_{(\beta)}^2)}\,,
\ee
whose derivation is in the appendix~\ref{app:KTNLambdasol}. The second concerns $\Phi_{(0)}$, the Myers-Perry scalar. In the zero-rotation case,
\be
\Phi_{(0)}\big|_{a_i=0}=\frac1{R^{D-3}}\,, \qquad \text{where}\qquad  R^2\equiv
\sum_{i=1}^{m}(x_i^2+y_i^2) + (1-\epsilon)z^2
\label{eq:cartR}
\ee
is the Cartesian spatial radius.
Resumming the small-rotation expansion, one can show that a Bessel function of the first kind arises:
\begin{equation}
\label{eq:Phi0J}
\Phi_{(0)} = 2^{\sigma}\,\Gamma(\sigma+1)\,
\frac{J_{\sigma}(\chi)}{\chi^{\sigma}}\,\frac{1}{R^{D-3}}\,,
\qquad
\sigma=\frac{D-5}{2}\,,
\qquad
\chi^{2} \equiv -\sum_{i} a_i^{2}\left(\partial_{x_i}^{2}+\partial_{y_i}^{2}\right)\,.
\end{equation}
Note that, in any $D$, $\frac{J_{\sigma}(\chi)}{\chi^{\sigma}}$ has a power-series expansion in $\chi^2$. This expression for $\Phi_{(0)}$ is the translation to position space of a momentum-space expression obtained in \cite{Bianchi:2024shc,Bianchi:2025xol}. We will obtain later its NUT counterpart. For this purpose, it will be useful to define the following `mass scalar' and `equal-NUT scalar': 
\be
\label{eq:PhiMPhiN}
\Phi_M\equiv\Phi_{(0)}\,, \qquad\qquad \Phi_N\equiv\sum_{\alpha=1}^{m-\epsilon} \,\Phi_{(\alpha)}\,.
\ee

\subsection{Double-copy structure}
\label{sec:KTNDC}

The double-copy interpretation of the Kerr-NUT metric is also conveniently expressed in Kerr-Schild coordinates \cite{Luna:2015paa,Chawla:2022ogv,Farnsworth:2023mff,Kokoska:2025voc}. The double copy (that is, the gravity solution) is \eqref{eq:mKS}:
\be
\label{eq:KTNmetricuv}
g_{\mu\nu}=\eta_{\mu\nu} + \sum_{\alpha=0}^{m-\epsilon}N_\alpha \,\Phi_{(\alpha)}\, \ell_{(\alpha)\mu}\ell_{(\alpha)\nu}\,.
\ee
The single copy (that is, the gauge theory solution), which following the literature we may call $\sqrt{\text{Kerr-NUT}}$, is
\be
\label{eq:A}
A_{\mu}= \sum_{\alpha=0}^{m-\epsilon}Q_\alpha \,\Phi_{(\alpha)}\, \ell_{(\alpha)\mu}\,,
\ee
where $Q_0$ is the electric charge and the $Q_{\alpha>0}$ are magnetic-type charges. This gauge field is a solution of the Maxwell equations on both the flat spacetime and the background \eqref{eq:KTNmetricuv}, as first observed in \cite{Krtous:2007xg}. Note that
\be
\nabla_\mu F^{\mu\nu} =\frac1{\sqrt{-g}}\partial_\mu (\sqrt{-g} F^{\mu\nu})= \partial_\mu ( \eta^{\mu\lambda}\eta^{\nu\rho} F_{\lambda\rho})= 0\,,
\ee
where in the second equality we used $\sqrt{-g}=1$ and $g^{\mu\lambda}g^{\nu\rho} F_{\lambda\rho}=\eta^{\mu\lambda}\eta^{\nu\rho} F_{\lambda\rho}$, the latter being a result of the non-trivial property $\ell_{(\alpha)}^\mu F_{\mu\nu}\propto \ell_{(\alpha)\nu}$. Finally, the zeroth copy (that is, the scalar solution) is
\be
\label{eq:phi}
\phi = \sum_{\alpha=0}^{m-\epsilon}c_\alpha \,\Phi_{(\alpha)}\,,
\ee
where the $c_\alpha$ are constant parameters. Using $\sqrt{-g}=1$ and $g^{\mu\nu}=\eta^{\mu\nu}-\sum_{\alpha=0}^{m-\epsilon}N_\alpha \,\Phi_{(\alpha)}\, \ell_{(\alpha)}^\mu\ell_{(\alpha)}^\nu$, the wave operator on the background \eqref{eq:KTNmetricuv} is
\be
\square\,\phi
=\frac1{\sqrt{-g}}\partial_\mu (\sqrt{-g}\,g^{\mu\nu} \partial_\nu \phi)
=
\square_{\rm flat}\,\phi
-
\sum_{\alpha=0}^{m-\epsilon}N_\alpha\,
\partial_\mu(
\Phi_{(\alpha)}
\ell_{(\alpha)}^\mu\ell_{(\alpha)}\cdot
\partial\phi
)\,.
\ee
The scalar \eqref{eq:phi} solves the wave equation on flat spacetime, $\square_{\text{flat}}\,\phi=0$, which fits the conventional double-copy picture, but not on the curved spacetime.

Building on earlier work \cite{Didenko:2008va}, ref.~\cite{Didenko:2022qxq} discussed a multi-copy structure for linearised spin-$s$ fields, which in our setting corresponds to considering
\be
\label{eq:psi}
\psi_{\mu_1\cdots\mu_s}=\sum_\alpha C_\alpha \Phi_{(\alpha)}\ell_{(\alpha)\mu_1}\cdots\ell_{(\alpha)\mu_s}\,.
\ee
We expect that these linearised spin-$s$ fields obey the obvious generalisations of our statements below on reality conditions and duality, in addition to obeying the Fronsdal spin-$s$ equation of motion. The Fronsdal equation in flat spacetime \cite{Fronsdal:1978} is
\be
\square_{\text{flat}}\, \psi_{\mu_1\cdots\mu_s}
-
s\,\partial_{(\mu_1}
\partial^\nu
\psi_{\mu_2\cdots\mu_s)\nu}
+
\frac{s(s-1)}{2}\,
\partial_{(\mu_1}\partial_{\mu_2}
\psi'_{\mu_3\cdots\mu_s)} =0\,,
\ee
where the trace\, $\psi'_{\mu_3\cdots\mu_s}
=
\eta^{\nu\rho}
\psi_{\nu\rho\mu_3\cdots\mu_s}$\, vanishes identically in our case. We note that, for any $s$, each NUT monomial is algebraically equivalent to the `mass' monomial already discussed in \cite{Didenko:2022qxq}, and therefore must satisfy the same equation of motion. Finally, we note that the naive covariantisation of the Fronsdal equation above is not gauge invariant on a generic curved background, including the Kerr-NUT background \eqref{eq:KTNmetricuv}; see \cite{Bekaert:2010hw} for a review of this topic. For recent work on the multi copy, including curvature/Weyl versions, see \cite{Brown:2025xlo,Misuna:2026has,Didenko:2026nip}.

\subsection{Reality conditions}
\label{sec:KTNreal}

The Kerr-NUT family was presented in \cite{Chen:2006xh} as a real $D$-dimensional solution with Lorentzian signature. Writing it in multi-Kerr-Schild form requires a complex coordinate transformation as described above (or a different metric signature as in \cite{Chen:2007fs}). The underlying reality of the Lorentzian spacetime requires that the quantities $N_\alpha \Phi_{(\alpha)}$ are real. Because the $\Phi_{(\alpha>0)}$ are imaginary in even $D$, it follows that the physical NUT charges are $i^{1-\epsilon}N_{\alpha>0}$. Considering also the physical normalisation of the charges, which we define as giving the ADM mass for $\alpha=0$, we have
\be
\label{eq:Nnorm}
N^\text{ADM}_\alpha = \frac{(D-2)\Omega_{D-2}}{16\pi G}\,N_\alpha \times \begin{cases} 
   i & \text{if $D$ even and $\alpha > 0$\,,} \\ 
   1 & \text{otherwise, } 
\end{cases} \qquad\quad \Omega_{D-2}= \frac{2\pi^{\frac{D-1}{2}}}{\Gamma\left(\frac{D-1}{2}\right)}\,.
\ee
This is not the only reality condition: the $\ell_{(\alpha)}$ are also complex if the corresponding $r_{(\alpha)}$ is, for any $D$. To illustrate how reality is achieved, it is useful to consider first the zeroth and single copies.

If the reality conditions on the $c_\alpha$ follow those of \eqref{eq:Nnorm}, then the quantities $c_\alpha \Phi_{(\alpha)}$ are real, and so is the scalar $\phi$ defined in \eqref{eq:phi}. Similarly, if the reality conditions on the $Q_\alpha$ follow those of \eqref{eq:Nnorm}, then the quantities $Q_\alpha \Phi_{(\alpha)}$ are real, but this is not sufficient to make $A_\mu$ defined in \eqref{eq:A} real, because the $\ell_{(\alpha>0)}$ are complex. Instead, one can show that
\be
\label{eq:realA}
d\; \text{Im($Q_\alpha \,\Phi_{(\alpha)}\, \ell_{(\alpha)}$)}=0\,, \quad \forall \alpha\,.
\ee
See the appendix~\ref{app:KTNLambdasol} for details. This means that $A_\mu$ is real up to a gauge transformation.

In gravity, as we mentioned, reality is guaranteed at the non-linear level by the coordinate transformations to match \cite{Chen:2006xh}, after imposing the reality conditions \eqref{eq:Nnorm}. Nevertheless, we can check reality at the linearised level, mirroring the gauge-theory case. We verified numerically in $D=4,5,6$ that
\be
\label{eq:realh}
R_{\mu\nu\lambda\rho}^{\text{linearised}}\left[\text{Im}(\Phi_{(\alpha)}\ell_{(\alpha)\delta}\ell_{(\alpha)\sigma})\right]=0\,, \quad \forall \alpha\,.
\ee
Hence, the metric perturbation Im($\Phi_{(\alpha)}\ell_{(\alpha)\mu}\ell_{(\alpha)\nu}$) is pure gauge.

\subsection{Odd $D$: trivial duality}
\label{sec:KTNoddD}

There are important differences between even and odd $D$, and we start by addressing the odd case. 
The relation \eqref{eq:idPhi} has a straightforward consequence in odd $D$, where the numerator is $1$ and the $\Phi_{(\alpha)}$ are all real. From the partial-fractions identity, it follows
\be
\label{eq:redunS}
\sum_{\alpha=0}^{m-1} \Phi_{(\alpha)}=0\,, \qquad \text{that is,} \qquad \Phi_M+\Phi_N = 0\,.
\ee
To understand the implications, we consider first the scalar $\phi$ defined in \eqref{eq:phi}. The identity above implies that one parameter among the $c_\alpha$ is redundant. We can map, say,
\be
\label{eq:redunc}
\{c_0,\ldots,c_{m-2},c_{m-1}\}\quad \mapsto \quad \{\hat c_0,\hat c_1,\ldots,\hat c_{m-1}\}=\{0,c_1-c_0,\ldots,c_{m-1}-c_0\}\,,
\ee
leaving $\phi$ unchanged. There is an analogous redundancy for the charges $Q_\alpha$ in the gauge field $A_\mu$ defined in \eqref{eq:A}: we find the closed 1-form relation
\be
\label{eq:redunA}
\sum_{\alpha=0}^{m-1} d\big(\Phi_{(\alpha)}\ell_{(\alpha)}\big)=0\,,
\ee
meaning that $\sum_{\alpha} \Phi_{(\alpha)}\ell_{(\alpha)}$ is pure gauge. See the appendix~\ref{app:KTNLambdasol} for details.

For gravity, a redundancy among the NUT charges was observed already in \cite{Chen:2006xh} for the odd-$D$ Kerr-NUT metric. This fits precisely with our observations for the zeroth and single copy. We checked numerically in $D=5,7$ the statement
\be
\label{eq:redunh}
\sum_{\alpha=0}^{m-1} R_{\mu\nu\lambda\rho}^{\text{linearised}}\left[\Phi_{(\alpha)}\ell_{(\alpha)\delta}\ell_{(\alpha)\sigma}\right]=0\,,
\ee
meaning that $\sum_{\alpha} \Phi_{(\alpha)}\ell_{(\alpha)\mu}\ell_{(\alpha)\nu}$ is a pure-gauge metric perturbation.
This is consistent with one of the charges $N_\alpha$ being redundant. We observe that this includes the mass, $N_0=M$, similarly to \eqref{eq:redunc}. Hence, in odd $D$, there is an essential ambiguity between the mass and (minus) the sum of the NUT charges.

Consider now the small-rotation resummation of $\Phi_M$ expressed in \eqref{eq:Phi0J} with a Bessel function $J_\sigma$.
We note that the mapping
\be
\label{eq:oddDdual}
\Phi_M \,\mapsto\, \Phi_N = - \Phi_M  \qquad \text{is reproduced by} \qquad J_\sigma  \,\mapsto\, (-1)^{\sigma+1} J_{-\sigma} = - J_\sigma\,,
\ee
where the last equality is a property of Bessel functions for integer $\sigma=(D-5)/2$, hence for odd $D$. So there is a duality exchanging the mass scalar and the equal-NUT scalar, but it is a trivial duality in odd $D$ because they differ only by a sign. In later sections, working in momentum space, we will discuss the analogous statements for gauge theory and gravity. We will also see that the duality becomes more interesting in even $D$.

Another feature is the simplicity of the equal-rotations limit, $a_i^2\to a^2$, especially for these scalars. For equal rotations, $\chi^2=- a^2\, \nabla^2_\text{spatial}$. Since $1/R^{D-3}$ is harmonic (for $R\neq0$), $\chi^2$ is effectively replaced by zero, so the rotation dependence drops out:
\be
\Phi_M\big|_{a_i^2=a^2}\, =\, \frac1{R^{D-3}}\, =\, - \Phi_N\big|_{a_i^2=a^2}\,,
\ee
for odd $D$.

\subsection{Even $D$: `electric-magnetic' duality}
\label{sec:KTNevenD}

The even-$D$ story is richer. Here, the mass scalar $\Phi_{M}$ is real, while the equal-NUT scalar $\Phi_{N}$ is imaginary, so $\Phi_M + \Phi_N \neq 0 $, in contrast with odd $D$. We have
\be
\Phi_M+\Phi_N = 
\sum_{\alpha=0}^{m}
\!\!\mathop{\mathrm{Res}}_{\;\;\;\;u=r_{(\alpha)}^2}  \frac{\sqrt{u}}{\prod_{\beta=0}^m
\big(u-r_{(\beta)}^2\big)}\,.
\ee
We choose the branch such that, in the patch $z>0$:
\be
r_{(0)}=+\sqrt{r_{(0)}^2}\,;
\qquad
r_{(\alpha)}=-i\sqrt{-r_{(\alpha)}^2}\,,
\quad
\alpha=1,\ldots,m \,.
\ee
For $z<0$, the signs are obtained by continuing the $r_{(\alpha)}$ as functions of $z$, rather than of $|z|$. 

Recall that $\Phi_M$ is given by \eqref{eq:Phi0J}. We find an analogous expression for $\Phi_N$, using the $D=4$ case and also the odd-$D$ case \eqref{eq:oddDdual} as clues. Putting the expressions for $\Phi_M$ and $\Phi_N$ together for comparison, we get in even $D$ 
\be
\label{eq:PhiJ}
\boldsymbol{\Big(}\Phi_{M},\Phi_N\boldsymbol{\Big)}
=
2^\sigma\Gamma(\sigma+1)\,
\frac{\boldsymbol{\Big(}J_\sigma(\chi),\,
i(-1)^{D/2}J_{-\sigma}(\chi)\boldsymbol{\Big)}}
{\chi^\sigma}
\frac1{R^{D-3}}\,,
\qquad
\sigma=\frac{D-5}{2}\,.
\ee
Hence,
\be
\label{eq:evenDdual}
\Phi_M \,\mapsto\, \Phi_N  \qquad \text{is reproduced by} \qquad J_\sigma  \,\mapsto\, i(-1)^{D/2} J_{-\sigma}\,.
\ee
This is analogous to \eqref{eq:oddDdual}; we note that $J_\sigma$ is independent of $J_{-\sigma}$ for half-integer $\sigma$.
Equivalently, in terms of a Hankel function of the first kind, we have
\be
\label{eq:PhiH}
\Phi_M+\Phi_N =
2^\sigma\Gamma(\sigma+1)\,
\frac{H_\sigma^{(1)}(\chi)}
{\chi^\sigma}
\frac1{R^{D-3}}= e^{i\chi} f^{(D)}(\chi)\, \frac1{R^{D-3}}\,,
\ee
where $f(\chi)$ is a polynomial in $1/\chi$, e.g.
\be
\label{eq:feqspin}
f^{(D=4)}(\chi)=1\,,\qquad f^{(D=6)}(\chi)=-\frac{i}{\chi}\,,\qquad f^{(D=8)}(\chi)=-\frac{3}{\chi^2}
\left(1+\frac{i}{\chi}\right)\,.
\ee
The real piece, $\Phi_M$, admits only even non-negative powers of $\chi$ acting on $1/R^{D-3}$. The imaginary piece, $\Phi_N$, admits only odd powers of $\chi$ acting on $1/R^{D-3}$; moreover, the lowest power is $\chi^{5-D}$, which is negative for $D\geq6$. The operation $\chi^{-1}$ is properly defined with recourse to Fourier space, where it corresponds to multiplying by $\xi^{-1}$, with $\xi= \sqrt{\sum_i a_i^2(k_{x_i}^2+k_{y_i}^2)}$. We will discuss Fourier space later when we investigate the scattering amplitudes. We verified our results for the scalars numerically in Fourier space up to $D=8$. Alongside numerical checks, we verified a version of our results analytically for equal rotations, $a_i^2= a^2$, as we discuss in the next section. Unlike the sum $\Phi_N$, the individual $\Phi_{(\alpha>0)}$ do not appear
to result purely from actions of $\chi$ or its inverse on $1/R^{D-3}$. This expectation is supported both by numerical explorations and by the
equal-rotations discussion below.

We claim that the duality described here for the scalars $\Phi_{M}$ and $\Phi_{N}$ extends to the Kerr-NUT solution, when we make the restriction that the NUT charges are equal, $N_{(\alpha>0)}=N$:
\be
\label{eq:mKSdual}
ds^2=ds^2_\text{flat} + M\,\Phi_{(0)}\, \ell_{(0)}^2 + N \sum_{\alpha=1}^{m-\epsilon} \,\Phi_{(\alpha)}\, \ell_{(\alpha)}^2\,.
\ee
Likewise for gauge theory, with equal monopole-like charges $Q_{(\alpha>0)}=Q_\text{mag}$:
\be
\label{eq:Adual}
A= Q_\text{el}\,\Phi_{(0)}\, \ell_{(0)} + Q_\text{mag} \sum_{\alpha=1}^{m-\epsilon} \,\Phi_{(\alpha)}\, \ell_{(\alpha)}\,,
\ee
where the first term is electric ($\sqrt{\text{Myers-Perry}}$), and the additional terms are magnetic. In $D=4$, this is the standard `electric-magnetic' duality. We will provide evidence for the duality in higher-$D$ gravity and gauge theory when we discuss the scattering amplitudes. Now, we recall that the traditional `electric-magnetic' duality is understood as a property of the linearised equations of motion. The conventional generalisation of the duality to higher $D$, based on the Hodge dual, involves fields of different tensor type \cite{Deser:1997mz,Hull:2001iu,Bunster:2013oaa}. In electromagnetism, it relates the 1-form photon field to a $(D-3)$-form field, known as the dual photon. In gravity, it relates the symmetric 2-tensor graviton to a mixed-symmetry tensor of Young type $(D-3,1)$, known as the dual graviton. What we have encountered here is a distinct generalisation of `electric-magnetic' duality, which relates ordinary photon/graviton fields to ordinary photon/graviton fields also in higher (even) $D$. It is realised in the particular stationary solutions we consider. As far as the authors understand it, however, it does not correspond to a symmetry of the linearised {\it dynamical} theory beyond $D=4$. We will return to this point when we discuss $2\!\mapsto\!2$ scattering. For completeness, we note that even in $D=4$ the duality does not hold generically in the non-linear theory \cite{Monteiro:2023dev}, though it holds in a sector of the solution space containing Kerr-NUT.

\subsection{Even $D$: equal-rotations case and the `little duality'}
\label{sec:KTNeqspin}

 We will now investigate the equal-rotations limit, $a_i^2\to a^2$, in even $D$. The simplicity of this limit allows us to write some very explicit expressions, but we will also find interesting subtleties. We will conclude with a comment on the Newman-Janis shift. This section is not essential for the main constructions presented in the paper. However, it will be useful for some later checks of these constructions.

In the equal-rotations case ($a_i^2=a^2, \forall i$), the spheroidal radial equation degenerates and has only 2 solutions for even $D$ (and 1 solution for odd $D$). Labelling the two solutions as $r_{(0)}^2$ and $r_{(1)}^2$, we have
\be
\label{eq:solseq}
r_{(0)}^2=u_+>0\,, \qquad  r_{(1)}^2=u_-<0\,,
\qquad
\text{with}
\qquad
u_\pm =\frac12\left(R^2-a^2\pm\sqrt{(R^2-a^2)^2+4a^2z^2}\right)\,.
\ee
For the other solutions of the generic-rotations case, we have $r_{(\alpha>1)}^2\!\to\!-a^2$ in the equal-rotations limit; see appendix~\ref{app:KTNLambdasoleq} for a proof. We will focus for now on $\Phi_{(0)}$ and $\Phi_{(1)}$, though we will comment later on the behaviour of the other scalars, $\Phi_{(\alpha>1)}$.
Consider the function $\mathcal I(r)=-\frac{iaz}{r}$. This is an involution, that is, $\mathcal I(\mathcal I(r))=r$. We choose the signs of $r_{(0)}$ and $r_{(1)}$ such that $r_{(0)}>0$ and
\be
r_{(1)} = \mathcal I(r_{(0)}) = - \frac{i a z}{r_{(0)}}\,.
\ee
In $D=4$, this involution exchanges the mass/electric and the NUT/magnetic charges, corresponding to the `electric-magnetic' duality. In higher $D$, this follows closely our previous generic-rotations story, with $\Phi_{(0)}=\Phi_{M}$, but $\Phi_{(1)}$ does not quite match $\Phi_{N}$, as we shall see. We will distinguish between the generic duality discussed before, and the `little duality' discussed here in the special equal-rotations case, which involves only $r_{(0)}$ and $r_{(1)}$. Explicitly,
\be
\Phi_{(0)}
=
\frac{r_{(0)}^3}
{(r_{(0)}^2+a^2)^{\frac{D-4}{2}}\,(r_{(0)}^4+a^2z^2)}
\,, 
\ee
and
\be
\label{eq:Phi1eq}
\Phi_{(1)}
=
\frac{r_{(1)}^3}
{(r_{(1)}^2+a^2)^{\frac{D-4}{2}}\,(r_{(1)}^4+a^2z^2)} =
\frac{i\,a z\, r_{(0)}\,(r_{(0)}^2+a^2)^{\frac{D-4}{2}}}
{\big(a^2(R^2 - z^2)\big)^{\frac{D-4}{2}}\,(r_{(0)}^4+a^2z^2)}\,.
\ee
We also have
\be
\label{eq:l0eq}
\ell_{(0)}
=
dt+
\frac{z}{r_{(0)}}\,dz
+
\frac{r_{(0)}}{r_{(0)}^2+a^2}
\sum_{i=1}^m(x_i dx_i+y_i dy_i)
+
\frac{a}{r_{(0)}^2+a^2}
\sum_{i=1}^m(x_i dy_i-y_i dx_i)
\,,
\ee
and
\be
\label{eq:lN}
\ell_{(1)}
=
dt
+
\frac{i r_{(0)}}{a}\,dz
-
\frac{i z r_{(0)}}{a(r_{(0)}^2-z^2)}
\sum_{i=1}^m(x_i dx_i+y_i dy_i)
+
\frac{r_{(0)}^2}{a(r_{(0)}^2-z^2)}
\sum_{i=1}^m(x_i dy_i-y_i dx_i)\,.
\ee

We encountered for generic rotations the expression \eqref{eq:PhiJ} expressing the duality. For the `little duality', we can similarly write
\be
\label{eq:PhieqJ}
\boldsymbol{\Big(}\Phi_{(0)},\Phi_{(1)}\boldsymbol{\Big)}
=
2^\sigma\Gamma(\sigma+1)\,
\frac{\boldsymbol{\Big(}J_\sigma(-a\partial_z),\,
i(-1)^{D/2}J_{-\sigma}(-a\partial_z)\boldsymbol{\Big)}}
{(-a\partial_z)^\sigma}
\frac1{R^{D-3}}\,,
\qquad
\sigma=\frac{D-5}{2}\,.
\ee
We have verified the expression for $\Phi_{(1)}$ by analytical and numerical computations in Fourier space up to $D=10$, and conjecture it to hold for any even $D$. Our branch choice identifies $\chi$ with $-a\partial_z$ for equal rotations, modulo an important caveat. Before the caveat, note that for equal rotations, $\chi^2=a^2 \big(\partial_z^2-\nabla^2_\text{spatial}\big)$, and this can effectively be replaced by $a^2\partial_z^2$ because $1/R^{D-3}$ is harmonic (for $R\neq0$), hence the relation between $\chi$ and $-a\partial_z$, with our choice of branch.

Now, let us be explicit about the difference between $\Phi_N$ and $\Phi_{(1)}$ for $D\geq6$. In $D=6$,
\be
\Phi_N= \Phi_{(1)} + \Phi_{(2)}\,.
\ee
Mirroring this equation term by term in the equal-rotations limit, we have
\be
-\frac{i\cos\chi}{\chi}\frac{1}{R^3}
=
-\frac{i\cos(-a\partial_z)}{-a\partial_z}\frac{1}{R^3}
-
\frac{i}{a\rho^2}\,, \qquad \text{where} \quad \rho^2=\sum_{i=1}^m(x_i^2+y_i^2) = R^2-z^2\,.
\ee
The term $\Phi_{(2)}=-\frac{i}{a\rho^2}$ captures the fact that the kernels of the operators $\chi$ and $-a\partial_z$ differ, so that the equal-rotations limit of $\chi^{-1}$ does not lead simply to $(-a\partial_z)^{-1}$. Indeed, $-\frac{i}{a\rho^2}$ is a zero mode of $\partial_z$. So what is the role of $\Phi_{(2)}$ here? We note that both $\Phi_{(1)}$ and $\Phi_{(2)}$ are singular at $\rho=0$; the most convenient expression to see this for $\Phi_{(1)}$ is \eqref{eq:Phi1eq}. On the other hand, the combination $\Phi_N$ does not share this singular locus. In $D\geq8$, the $\Phi_{(\alpha>1)}$ individually diverge in the equal-rotations limit; their sum is finite at a generic position, but is singular at $\rho=0$, precisely in such a way as to cancel the singular behaviour of $\Phi_{(1)}$ on that locus, so that $\Phi_N$ is finite. To be concrete, in even $D$ dimensions,
\be
\Phi_N= \Phi_{(1)} + \sum_{\alpha=2}^{m} \,\Phi_{(\alpha)}\,,
\ee
with $m = (D-2)/2$.
In the equal-rotations limit, we have
\be
\prod_{\alpha=0}^{m}(u-r_{(\alpha)}^2) = (u+a^2)^{m-1}Q(u)\,, \qquad Q(u) =u(u+a^2)-\rho^2 u-z^2(u+a^2) = (u-u_+)(u-u_-)\,,
\ee
where $u_\pm$ were defined in \eqref{eq:solseq}. Then, we find
\be
\label{eq:Phiresteq}
\sum_{\alpha=2}^{m} \,\Phi_{(\alpha)} = \operatorname*{Res}_{u=-a^2}
\frac{-i\sqrt{-u}\,du}
{(u+a^2)^{m-1}Q(u)} =
\frac{1}{(m-2)!}
\left.
\frac{d^{m-2}}{du^{m-2}}
\left(
\frac{-i\sqrt{-u}}{Q(u)}
\right)
\right|_{u=-a^2}\,.
\ee
For $D=6$, we obtain $-\frac{i}{a\rho^2}$ as seen earlier. For $D\geq8$, the resulting expression depends on $z$ polynomially with degree $D-6$, hence\, $\sum_{\alpha=2}^{m} \,\Phi_{(\alpha)}\in \text{ker}\, \partial_z^{D-5}$\,. We note that $\partial_z^{-(D-5)}$ arises in \eqref{eq:PhieqJ}, so what $\Phi_{(1)}$ misses with respect to $\Phi_N$ are the zero-modes of $\partial_z^{D-5}$. Another remark is that, similarly to $\Phi_{(1)}$, the expression is always singular at $\rho=0$ because $Q(-a^2)=a^2\rho^2$. Hence, these contributions do not decay in all directions for large $R$, so they are not `localised'. We find, in the limit $|z|\to\infty$ at fixed $\rho$, 
\[
\Phi_{(1)}
\sim
+\frac{i z^{2m-4}}{a^{2m-3}\rho^{2m-2}}
+\cdots,
\qquad
\sum_{\alpha=2}^{m} \,\Phi_{(\alpha)}
\sim
-\frac{i z^{2m-4}}{a^{2m-3}\rho^{2m-2}}
+\cdots \,,
\]
which illustrates how this `unlocalised' behaviour cancels in $\Phi_N$. This notion of `localised' versus `unlocalised' is also reflected in the distributional support of the action of the Laplacian $\nabla^2_\text{spatial}$ on these scalars. Now, with the expression \eqref{eq:Phiresteq} in hand, and with the expression for $\Phi_{(1)}$ in \eqref{eq:Phi1eq}, we obtain a relatively simple explicit expression for $\Phi_N$ in the even-$D$ equal-rotations limit,
\be
\Phi_N = \frac{r_{(1)}^3}
{(r_{(1)}^2+a^2)^{m-1}\,(r_{(1)}^4+a^2z^2)} + \frac{1}{(m-2)!}
\left.
\frac{d^{m-2}}{du^{m-2}}
\left(
\frac{-i\sqrt{-u}}{Q(u)}
\right)
\right|_{u=-a^2}\,.
\ee

For clarity, the complete `electric-magnetic' dual of $\Phi_M=\Phi_{(0)}$ is $\Phi_N$, as stated in \eqref{eq:evenDdual} for generic rotations. However, the symmetry of the equal-rotations case allows for the interpretation of $\Phi_{(1)}$ as a simpler (if unlocalised) `little dual' of $\Phi_{(0)}$.

Finally, let us make a comment on the Newman-Janis shift \cite{Newman:1965tw}. In $D=4$, the Kerr metric can be obtained from the Schwarzschild metric via a complex shift. This almost generalises in the higher-even-$D$ equal-rotations sector. 
We can rewrite \eqref{eq:PhieqJ} in a manner that mirrors \eqref{eq:PhiH}:
\be
\label{eq:Phi01}
\Phi_{(0)}+\Phi_{(1)}
=
e^{-ia\partial_z}
f^{(D)}(-a\partial_z)\,
\frac{1}{R^{D-3}}
= \left(f^{(D)}(-a\partial_z)\,
\frac{1}{R^{D-3}}\right) \Bigg|_{z\to z-ia}
\,,
\ee
where $f^{(D)}(-a\partial_z)\,
\frac{1}{R^{D-3}}$ evaluates to
\be
\frac1{R} \quad \text{in} \;\; D=4\,, \qquad 
\frac{i}{a}\,\partial_z^{-1}\frac{1}{R^3} =\frac{iz}{a\rho^2R} \quad \text{in} \;\; D=6\,,
\ee
and
\be
\left(-\frac{3}{a^2}\,\partial_z^{-2} +
\frac{3i}{a^3}\,\partial_z^{-3}\right)\frac{1}{R^5}=
-\frac{R^2+z^2}{a^2\rho^4R} + \frac{i zR}{a^3\rho^4}
\quad \text{in} \;\; D=8\,.
\ee
We also note that \,$R\big|_{z\to z-ia} = r_{(0)}+r_{(1)} $\,.
As we can see, in $D=4$, the `rotating Coulombic' function $\Phi_{(0)}$ is the real part of the complex shift of the `static Coulombic' function $1/R$. There is also a procedure to `add rotation' to the Schwarzschild 1-form, $\ell_{(0)}|_{a=0}$, leading to the Kerr 1-form, $\ell_{(0)}$, given by \eqref{eq:l0eq} with $m=1$. For higher $D$, however, we see that the complex shift is not sufficient already for $\Phi_{(0)}$, due to the action of $f^{(D)}(-a\partial_z)$.


\section{Kerr-NUT scattering amplitudes}
\label{sec:amps}

\subsection{Metrics $\leftrightarrow$ amplitudes}
\label{sec:metricsamps}

The black hole metrics define 3-point scattering amplitudes on flat spacetime, and vice-versa, as follows. In de Donder gauge (to stay generic for now, as not all spacetimes admit a Kerr-Schild-type gauge), the linearised metric perturbation caused by an object with velocity $u^\mu$ and energy-momentum tensor $T_{\mu \nu }$ obeys
\be
\label{eq:h}
h_{\mu \nu }(x)=\frac{\kappa^2}{2} \int \frac{d^{D}k}{(2\pi )^{D}}\,\frac{e^{ik\cdot x}}{k^{2}}\,\delta(k\cdot u)\,\bar{T}_{\mu \nu }(k)\,,
\qquad
\bar{T}_{\mu\nu} \equiv T_{\mu \nu }-\frac{1}{D-2}\,\eta_{\mu \nu }\,T^\lambda_\lambda\,,
\ee
where $\, g_{\mu \nu }=\eta_{\mu \nu }+ h_{\mu \nu }$\,. Our convention for the Einstein equations is $R_{\mu\nu}-\frac1{2}g_{\mu\nu}R=8\pi GT_{\mu\nu}$, with $\kappa^2=32\pi G$, and we use the mostly-plus metric signature as before. We define the scattering amplitude as
\be
{\mathcal A}_\text{grav}(\varepsilon,k,u) = \frac{\kappa^2}{2}\, \varepsilon^{\mu \nu }(k)\, \bar{T}_{\mu \nu }(k)\,, \quad \text{supported on $k^2=0$,}
\label{eq:Amp}
\ee
with a normalisation that will prove convenient; the canonically normalised amplitude is ${\mathcal A}_\text{grav}/\kappa$.
Note that, already in the expression \eqref{eq:h}, only the piece of $\bar{T}_{\mu \nu }(k)$ supported on $k^2=0$ contributes: if we change $\bar{T}_{\mu \nu }(k)$ by a term $k^{2}\times$(analytic in $k$), then that term does not contribute to ${h}_{\mu \nu }(x)$ outside of the object's `worldline'. The gauge-invariant version of \eqref{eq:h} is the formula for the linearised curvature
\be
\label{eq:R}
R_{\mu \nu }{}^{\lambda\rho}(x)=\kappa^2 \int \frac{d^{D}k}{(2\pi )^{D}}\,\frac{e^{ik\cdot x}}{k^{2}}\,\delta(k\cdot u)\, k_{[\mu}k^{[\lambda}\bar{T}_{\nu]}{}^{\rho]}\,.
\ee
To connect to the scattering amplitude \eqref{eq:Amp}, we can rewrite this as
\be
\label{eq:RA}
R_{\mu \nu }{}^{\lambda\rho}(x)=2 \sum_{\eta} \int \frac{d^{D}k}{(2\pi )^{D}}\,\frac{e^{ik\cdot x}}{k^{2}}\,\delta(k\cdot u)\, k_{[\mu}k^{[\lambda}{\varepsilon^{({\eta})}}_{\nu]}{}^{\rho]}\,{\mathcal A}_\text{grav}(\varepsilon^{({\eta})},k,u)\,, 
\ee
where $\eta$ labels the $D(D-3)/2$ polarisations of the graviton. This expression matches \eqref{eq:R} after using the completeness relation
\be
\sum_{\eta} \varepsilon^{(\eta)}_{\mu\nu}\, \varepsilon^{(\eta)}_{\alpha\beta}
= \frac{1}{2} \bigl(  P_{\mu\alpha}  P_{\nu\beta} +  P_{\mu\beta}  P_{\nu\alpha} \bigr)
- \frac{1}{D-2}\,  P_{\mu\nu}  P_{\alpha\beta} \,, \qquad 
 P_{\mu\nu} \equiv \eta_{\mu\nu} - \frac{k_\mu n_\nu + n_\mu k_\nu}{k\cdot n}\,,
\ee
where $n$ is a null reference vector such that $\varepsilon^{(\eta)}_{\mu\nu} n^\nu=0$. The dependence on $n$ drops out of \eqref{eq:RA} due to gauge invariance. A brief discussion of the completeness relation, and its extension for any spin, is included in appendix~\ref{app:projector}.

The expression \eqref{eq:RA} relating the curvature and the 3-point amplitude can also be obtained from a quantum-first on-shell approach using the KMOC formalism \cite{Kosower:2018adc,Cristofoli:2021vyo} as presented in \cite{Monteiro:2020plf}. To be precise, the on-shell conditions for 3-point amplitudes have no support in real Lorentzian kinematics, so writing a Fourier integral that is explicitly on-shell --- featuring $\delta(k^2)$ in the integrand --- requires an analytic continuation, which in \cite{Monteiro:2020plf} was enacted by considering split signature. Once a split-signature solution is obtained, one may continue back to Lorentzian signature. Modulo this technicality, our discussion above is equivalent to the on-shell prescription.

While the amplitude appears to correspond only to the linearised metric, the fact that we are dealing with multi-Kerr-Schild spacetimes means that this correspondence is exact.


\subsection{Mass/electric (Myers-Perry) amplitudes}
\label{sec:MPamps}

The 3-point scattering amplitudes generating Myers-Perry black holes were recently obtained \cite{Bianchi:2024shc,Bianchi:2025xol}; see also \cite{Gambino:2024uge,Gambino:2025iyx,Akpinar:2025huz,Campanella:2026wqt,Cangemi:2026cjy}. The expression for $T_{\mu\nu}(k)$, which determines the amplitude via \eqref{eq:Amp}, was found in \cite{Bianchi:2024shc} to be

\begin{align}\label{eq:Tmunu_norm}
T_{\mu\nu}(k) = 
\frac{
2^{\frac{D-1}{2}}\pi\,
\Gamma\!\left(\frac{D-1}{2}\right)
}{D-2}\, M^\text{ADM}
\Bigg[&\,
u_\mu u_\nu\,F_1^{(D)}(\xi)
+\frac{F_2^{(D)}(\xi)}{\xi^2}\,(S\cdot k)_\mu (S\cdot k)_\nu \nonumber \\
&\quad -i\big(u_\mu (S\cdot k)_\nu + u_\nu (S\cdot k)_\mu\big)\,F_3^{(D)}(\xi) \Bigg]\,,
\end{align}
where in the prefactor $M^\text{ADM}=N^\text{ADM}_0$, with the normalisation \eqref{eq:Nnorm}.
Moreover, $u$ is the velocity, which obeys $u\cdot u=-1$, and $J_{\mu\nu}$ is the angular momentum tensor, which obeys $J_{\mu\nu}u^\mu=0$. In the frame of the black hole, $u^\mu=\delta^\mu_t$, and the non-vanishing components of $J_{\mu\nu}$ are $J_{x_iy_i}=-J_{y_ix_i}$, for $i=1,\cdots, m$\,, where we employ the coordinates in \eqref{eq:ds2flat}. We define
\be
\label{eq:Suv}
S_{\mu\nu}\equiv\frac{D-2}{2}\, \frac{J_{\mu\nu}}{M^\text{ADM}}\,, \qquad \quad a_i = S_{x_iy_i}\,,
\ee
where the $a_i$ are the Myers-Perry rotation parameters. The quantity $\xi$ in \eqref{eq:Tmunu_norm} is
\be
\xi \equiv \sqrt{(S\cdot k)^2}
\;=\; \sqrt{\sum_i a_i^2(k_{x_i}^2+k_{y_i}^2)}\,.\label{eq:zeta_def}
\ee
Finally,
\begin{equation}\label{eq:Zn}
\begin{aligned}
F_2^{(D)}(\xi) &= -\xi\,\mathcal{Z}_1^{(D)}(\xi),\\
F_3^{(D)}(\xi) &= \mathcal{Z}_0^{(D)}(\xi),\\
F_1^{(D)}(\xi) &= F_2^{(D)}(\xi) + (D-2)\,F_3^{(D)}(\xi)\,,
\end{aligned}
\qquad \quad \mathcal{Z}_n^{(D)}(\xi) \;\equiv\; \frac{J_{n+\frac{D-3}{2}}(\xi)}{\xi^{\frac{D-3}{2}}}\,,
\end{equation}
A note on the relation of our conventions to refs.~\cite{Bianchi:2024shc,Bianchi:2025xol}: our $\xi$ matches the quantity $u_a$ of \cite{Bianchi:2025xol}, and the quantity $\frac{D-2}{2}\,\zeta$ of \cite{Bianchi:2024shc}, where the latter normalises the $a_i$ differently; their $d$ is $D-1$; we have $-i$ instead of $i$ in \eqref{eq:Tmunu_norm} due to our convention for the Fourier transform; our $T^{\mu\nu}$ has an additional $2\pi$ factor with respect to the analogous quantity in \cite{Bianchi:2024shc} as they effectively have $2\pi\delta(k_t)$ instead of our $\delta(u\cdot k)$ in \eqref{eq:h}.

We will now rewrite this result from \cite{Bianchi:2024shc,Bianchi:2025xol} for the Myers-Perry source. Starting with the scalar, and using our language from the previous sections,
\be
\Phi_M(x) \equiv \Phi_{(0)}(x)
=
\int\frac{d^Dk}{(2\pi)^D}
\frac{e^{ik\cdot x}}{k^2}\,
\delta(k\cdot u)\,
\mathcal A_\text{scalar}^M(k,u) \,,
\ee
where the scalar amplitude is
\be
\mathcal A_\text{scalar}^M(k,u) = (2\pi)^{\frac{D+1}{2}}\,
\frac{J_\frac{D-5}{2}(\xi)}{\xi^\frac{D-5}{2}}\,.
\ee
For the gauge field and gravity, it is convenient to introduce the Bessel order-shifting differential operator
\be
\label{eq:Dxi}
\mathscr D_\xi\equiv\frac{1}{\xi}\frac{d}{d\xi}\,,
\qquad \text{such that} \qquad
\mathscr D_\xi^n
\left(
\frac{J_{\pm\rho}(\xi)}{\xi^\rho}
\right)
=
(\mp1)^n
\frac{J_{\pm(\rho+ n)}(\xi)}{\xi^{\rho+n}}\,.
\ee
Then, we have for the gauge field, in particular for the electric monomial,
\be
\label{eq:AMfourier}
A^{M}_\mu(x) \equiv \Phi_{(0)}\, {\ell_{(0)}}_\mu = 
\int\frac{d^Dk}{(2\pi)^D}
\frac{e^{ik\cdot x}}{k^2}\,
\delta(k\cdot u)\,
\mathcal J^M_\mu(k,u) \,,
\ee
where, up to a gauge transformation ($\propto k_\mu$),
\be
\mathcal J^M_\mu(k,u)
=
\left[
-u_\mu
-i(S\cdot k)_\mu\,\mathscr D_\xi
\right] \mathcal A_\text{scalar}^M(k,u)\,,
\qquad
(S\cdot k)_\mu=S_{\mu\nu}k^\nu\, .
\ee
We can also write the field strength in terms of the scattering amplitude generating the solution,
\be
\label{eq:FA}
{F}_{\mu\nu}(x)
=
2i\sum_{\eta}
\int\frac{d^Dk}{(2\pi)^D}
\frac{e^{ik\cdot x}}{k^2}\,
\delta(k\cdot u)\,
k_{[\mu}{\epsilon^{(\eta)}}_{\nu]}(k)\,
\mathcal A_{\rm gauge}(\epsilon^{(\eta)},k,u)\,.
\ee
For the electric monomial \eqref{eq:AMfourier}, the amplitude is
\be
\label{eq:MampA}
\mathcal A^M_{\rm gauge}(\epsilon,k,u)
=
\epsilon^\mu(k)\,\mathcal J^M_\mu(k,u)=
\left[
-\epsilon\cdot u
-i\,\epsilon\cdot S\cdot k\,\mathscr D_\xi
\right] \mathcal A_\text{scalar}^M(k,u)\,,
\ee
which is manifestly gauge-invariant subject to $k\cdot u=0$. This can be seen as a spin-raising operator acting on the scalar amplitude, which we will exploit in a later section.

Turning now to gravity, the mass (Myers-Perry) monomial is
\be
\label{eq:hM}
h^M_{\mu\nu}(x) \equiv \Phi_{(0)}\,{\ell_{(0)}}_\mu\,{\ell_{(0)}}_\nu
=
\int\frac{d^Dk}{(2\pi)^D}
\frac{e^{ik\cdot x}}{k^2}\,
\delta(k\cdot u)\,
\mathcal J^M_{\mu\nu}(k,u)\,.
\ee
We note that $\mathcal J^M_{\mu\nu}$ is not proportional to $\bar T_{\mu\nu}$ obtained from \eqref{eq:Tmunu_norm}, because the two expressions are written in different gauges (Kerr-Schild versus de Donder). However, their contractions with an on-shell transverse-traceless graviton polarisation tensor agree. We can write
\begin{align}
\label{eq:Mamph}
\mathcal A^M_{\rm grav}(\varepsilon,k,u)
&=
\varepsilon^{\mu\nu}(k)\,
\mathcal J^M_{\mu\nu}(k,u) \nonumber\\
&=
\Bigg[
\varepsilon_{\mu\nu}u^\mu u^\nu
\left(
1-\mathscr D_\xi
\right)
+
2i\,\varepsilon_{\mu\nu}u^\mu(S\cdot k)^\nu\mathscr D_\xi
-
\varepsilon_{\mu\nu}(S\cdot k)^\mu(S\cdot k)^\nu
\mathscr D_\xi^2
\Bigg]
\mathcal A^M_{\rm scalar}(k,u)\,,
\end{align}
where we dropped terms where $\varepsilon^{\mu\nu}$ contracts with $k_\mu$, $k_\nu$, or $\eta_{\mu\nu}$.
This expression is consistent with \eqref{eq:Tmunu_norm}, with the normalisation  
\be
\mathcal A_{\rm grav}(\varepsilon,k,u) \Big|_\text{Myers-Perry}
=
M\,
\mathcal A^M_{\rm grav}(\varepsilon,k,u)\,,
\ee
where $M=N_0$ is the geometric mass parameter. To check the equivalence to \eqref{eq:Tmunu_norm}, it is convenient to use the Bessel identity
\be
\label{eq:Besselid}
\big(\xi^2 \mathscr D_\xi^2 +2(\rho+1)\mathscr D_\xi + 1\big) \frac{J_\rho(\xi)}{\xi^\rho} =0\,.
\ee


\subsection{Equal-NUT/magnetic amplitudes from duality}
\label{sec:KNamps}

We now turn to the amplitudes for the dual solutions, namely those with equal NUT/magnetic charges. The scalar amplitude is obtained by flipping the Bessel order,
\be
\label{eq:A0N}
\mathcal A_{\rm scalar}^N(k,u)
=
\eta_D\,(2\pi)^{\frac{D+1}{2}}\,
\frac{J_{-\frac{D-5}{2}}(\xi)}{\xi^{\frac{D-5}{2}}}\,, 
\qquad 
\eta_D = \begin{cases} i(-1)^{\frac{D}{2}}\,, & D \ \text{even},\\[2mm] (-1)^{\frac{D-3}{2}}\,, & D \ \text{odd}, \end{cases}
\ee
which follows from \eqref{eq:oddDdual}/\eqref{eq:evenDdual} for odd/even $D$.

Because the amplitudes \eqref{eq:MampA} and \eqref{eq:Mamph} for gauge theory and gravity in the mass/electric sector are obtained by a spin-raising operation on a seed scalar amplitude, one may hope that the same happens in the equal-NUT/magnetic sector. So the natural guesses are
\be
\label{eq:NampA}
\mathcal A^N_{\rm gauge}(\epsilon,k,u)
=
\left[
-\epsilon\cdot u
-i\,\epsilon\cdot S\cdot k\,\mathscr D_\xi
\right] \mathcal A_\text{scalar}^N(k,u)
\ee
and
\be
\label{eq:Namph}
\mathcal A^N_{\rm grav}(\varepsilon,k,u)
=
\Bigg[
\varepsilon_{\mu\nu}u^\mu u^\nu
\left(
1-\mathscr D_\xi
\right)
+
2i\,\varepsilon_{\mu\nu}u^\mu(S\cdot k)^\nu\mathscr D_\xi
-
\varepsilon_{\mu\nu}(S\cdot k)^\mu(S\cdot k)^\nu
\mathscr D_\xi^2
\Bigg]
\mathcal A^N_{\rm scalar}(k,u)\,. 
\ee
It turns out that these guesses are correct, which is one of the main results of this paper. The validity of the formulas relies crucially on our choice of representation of the spin-raising operation, given the ambiguity introduced by the Bessel identity \eqref{eq:Besselid}. In particular, we fixed the ambiguity for the spin-2 operation by demanding that the coefficients are independent of $D$. It turns out that this coincides with the possibility of writing the spin-2 operation as the doubled action of a spin-raising operator. We will discuss in later sections this structure and the tests we made to verify the formulas above for the $N$-sector amplitudes.

We note that
\be
\label{eq:oddDdualAmps}
\text{in odd $D$,} \qquad  \mathcal A^N_{\rm scalar/gauge/grav} = - \mathcal A^M_{\rm scalar/gauge/grav}\,.
\ee
The cases of gauge theory and gravity follow the seed scalar case, via the spin-raising operations. In turn, the scalar case is a direct translation to momentum space of our discussion of parameter redundancy in section~\ref{sec:KTNoddD}. So the trivial duality discussed there for odd $D$ extends to the theories of spin 1 and spin 2.

We find
\be
\label{eq:evenDdualAmps}
\text{in even $D$,} \qquad  \mathcal A^N_{\rm scalar/gauge/grav} = i(-1)^{D/2}\, \mathcal A^M_{\rm scalar/gauge/grav}\Big|_{J_{\frac{D-5}{2}}\,\mapsto \,J_{-\frac{D-5}{2}}}\,,
\ee
where the Bessel function replacement is applied only to the scalar seed, that is, {\it before} the spin-raising operations. Revisiting the Bessel order-shifting formula
\be
\mathscr D_\xi^n
\left(
\frac{J_{\pm\rho}(\xi)}{\xi^\rho}
\right)
=
(\mp1)^n
\frac{J_{\pm(\rho+ n)}(\xi)}{\xi^{\rho+n}}\,,
\ee 
we now see that the $+\rho$ and $-\rho$ cases are relevant to the $M$ and $N$ sectors, respectively. For clarity, the amplitudes generating the even-$D$ solutions \eqref{eq:Adual} and \eqref{eq:mKSdual} are
\be
\label{eq:AgaugeAgrav}
\mathcal A_{\rm gauge} = Q_\text{el} \,\mathcal A^M_{\rm gauge} + Q_\text{mag}\, \mathcal A^N_{\rm gauge}
\qquad \text{and} \qquad 
\mathcal A_{\rm grav} = 
M\, \mathcal A^M_{\rm grav} + N\,\mathcal A^N_{\rm grav}\,.
\ee


\subsection{Simplifications for $D=4$}
\label{sec:4Damps}

We will briefly describe the simplifications that occur in $D=4$, to match the literature on nutty/dyonic amplitudes \cite{Huang:2019cja,Emond:2020lwi}. In $D=4$, it is convenient to use a helicity basis of polarisation vectors, $\epsilon^\pm_\mu$. Under $k^2=0$ and $k\cdot u=0$, we choose the basis such that
\be
\epsilon^\pm\cdot \frac{(S\cdot k)}{\xi} = \mp\, \epsilon^\pm \cdot u\,,
\qquad \text{where} \qquad \xi = i a\cdot k\,.
\ee
Here, \,$a^\mu = -\frac12\, \epsilon^{\mu\nu\rho\sigma} u_\nu S_{\rho\sigma}$\, is the rotation vector; our convention is $\epsilon^{0123}=+1$.
We note that $\xi$ is the quantity from previous sections: in the rest frame, where $u^\mu=\delta^\mu_t$, we have $\xi^2=a^2(k_x^2+k_y^2)$, so under the kinematic constraints we obtain $\xi^2=-a^2k_z^2$.

The gauge-theory amplitude in \eqref{eq:AgaugeAgrav} becomes
\begin{align}
\mathcal A_{\rm gauge}(\epsilon^\pm)
& =
-(\epsilon^\pm\cdot u) \left[
1
\mp i\,\frac{d}{d\xi}
\right]\Big(Q_\text{el}\,\mathcal A^M_{\rm scalar}+Q_\text{mag}\,\mathcal A^N_{\rm scalar}\Big) \nonumber\\
& = - 8\pi^2\,(\epsilon^\pm\cdot u) \left[
1
\mp i\,\frac{d}{d\xi}
\right]\Big(Q_\text{el}\,\cos\xi+i\,Q_\text{mag}\,\sin\xi\Big)\nonumber\\
& = - 8\pi^2\,(Q_\text{el}\pm Q_\text{mag})\,(\epsilon^\pm\cdot u)\, e^{\mp a\cdot k}
\,.
\end{align}
Recall our reality conditions from section~\ref{sec:KTNreal}: in even $D$, our geometric parameter $Q_\text{mag}$ is $i$ times the physical (real) magnetic charge. Hence, we can write $Q_\text{el}\pm Q_\text{mag}=e^{\pm i\theta}\sqrt{Q_\text{el}^2-Q_\text{mag}^2}$, where $\theta$ is the dyonic phase. With this note, we match \cite{Huang:2019cja,Emond:2020lwi} up to normalisation.

Similarly, the gravity amplitude in \eqref{eq:AgaugeAgrav}, written in a basis of polarisation tensors $\varepsilon^\pm_{\mu\nu}=\epsilon^\pm_\mu\epsilon^\pm_\nu$, becomes
\be
\mathcal A_{\rm grav}(\varepsilon^\pm)
=
16\pi^2\,
(\epsilon^\pm\cdot u)^2
\left(M\pm N\right)
e^{\mp a\cdot k}\,.
\ee
Given the even-$D$ reality condition that our geometric parameter $N$ is $i$ times the physical NUT charge, we again match \cite{Huang:2019cja,Emond:2020lwi}  up to normalisation.


\subsection{Higher-spin amplitudes from a spin-raising operator}
\label{sec:HSamps}

We wrote down amplitudes for the $M$ and $N$ parts of scalar (spin 0), gauge theory (spin 1) and gravity (spin 2). In this section, we will extend this construction to spin $s$. Our starting point in position space is the multi-Kerr-Schild form \eqref{eq:psi}. We need the de Wit-Freedman higher-spin generalisation of the curvature \cite{deWit:1979sib}:
\be
\label{eq:hsR}
\mathcal R^{\text{spin-}s}_{\mu_1\nu_1|\cdots|\mu_s\nu_s}
\equiv
2^{s}\,
\partial_{[\mu_1}\cdots\partial_{[\mu_s}
\psi_{\nu_1]\cdots\nu_s]}\,,
\ee
where the anti-symmetrisation is performed separately in each pair
$(\mu_a,\nu_a)$. With this normalisation, we have
\be
\mathcal R^{\text{spin-}1}_{\mu_1\nu_1} = F_{\mu_1\nu_1}\,, \qquad
\mathcal R^{\text{spin-}2}_{\mu_1\nu_1|\mu_2\nu_2} = -2 \,R_{\mu_1\nu_1\mu_2\nu_2}\,,
\ee
where $R_{\mu_1\nu_1\mu_2\nu_2}$ is the linearised Riemann curvature. We need to translate between position space, in terms of the linearised curvature, and momentum space, in terms of the scattering amplitude. Generalising \eqref{eq:FA} for spin 1, and \eqref{eq:RA} for spin 2, the map is
\begin{equation}
\label{eq:hsRA}
\mathcal R^{\text{spin-}s}_{\mu_1\nu_1|\cdots|\mu_s\nu_s}(x)
= (2i)^s
\sum_{\eta}
\int
\frac{d^D k}{(2\pi)^D}\,
\frac{e^{ik\cdot x}}{k^2}\,
\delta(k\cdot u)\,
k_{[\mu_1}\cdots k_{[\mu_s}
\varepsilon^{(\eta)}_{\nu_1]\cdots\nu_s]}(k)
\,\mathcal A_{\text{spin-}s}(\varepsilon^{(\eta)},k,u)\,,
\end{equation}
where the anti-symmetrisation is again performed separately in each pair $(\mu_a,\nu_a)$. Similarly to spin 1 and 2, the sum over polarisations
\be
\sum_{\eta} \varepsilon^{(\eta)}_{\nu_1\cdots\nu_s}\, \varepsilon^{(\eta)}_{\alpha_1\cdots\alpha_s}
\ee
is dealt with via the completeness relation, which we describe in appendix~\ref{app:projector}.

We will present expressions for $\mathcal A_{\text{spin-}s}^M$ and $\mathcal A_{\text{spin-}s}^N$ defined as the amplitudes associated to the fields
\be
\psi_{\mu_1\cdots\mu_s}^M \equiv \Phi_{(0)}\,
\ell_{(0)\mu_1}\cdots \ell_{(0)\mu_s} \qquad \text{and} \qquad
\psi_{\mu_1\cdots\mu_s}^N \equiv \sum_{\alpha=1}^{m-\epsilon} \Phi_{(\alpha)}\,
\ell_{(\alpha)\mu_1}\cdots \ell_{(\alpha)\mu_s}\,,
\ee
respectively.

For odd $D$ ($\epsilon=1$), we expect
\be
\psi_{\mu_1\cdots\mu_s}^M +
\psi_{\mu_1\cdots\mu_s}^N = \partial_{(\mu_1}
\lambda_{\mu_2\cdots\mu_s)}\,,
\ee
which is a pure-gauge field. This generalises the pattern up to spin 2, discussed in section~\ref{sec:KTNoddD}. The amplitudes statement, which generalises \eqref{eq:oddDdualAmps}, is
\be
\label{eq:HSoddDdualAmps}
\text{in odd $D$,} \qquad  \mathcal A^N_{\text{spin-}s} = - \mathcal A^M_{\text{spin-}s}\,.
\ee

For even $D$ ($\epsilon=0$), the expectation is that there is a non-trivial duality between $\psi_{\mu_1\cdots\mu_s}^M$ and $\psi_{\mu_1\cdots\mu_s}^N$, again following the pattern seen up to spin 2. The natural generalisation of \eqref{eq:evenDdualAmps} is
\be
\label{eq:HSevenDdualAmps}
\text{in even $D$,} \qquad  \mathcal A^N_{\text{spin-}s} = i(-1)^{D/2}\, \mathcal A^M_{\text{spin-}s}\Big|_{J_{\frac{D-5}{2}}\,\mapsto \,J_{-\frac{D-5}{2}}}\,,
\ee
where again the Bessel order reflection acts only on the scalar seed amplitude. We will now make this explicit.

We want to construct expressions for these scattering amplitudes. We already know the cases $s=0,1,2$. In particular, we can write \eqref{eq:MampA} and \eqref{eq:NampA} as
\be
\mathcal A^X_{\text{spin-}1}(\epsilon,k,u)
=
\epsilon^\mu\,\mathcal S_\mu\,
 \mathcal A_{\text{spin-}0}^X(k,u)\,, 
 \qquad X=M,N\,,
\ee
using the spin-raising operator
\begin{equation}
\label{eq:spinraising}
\mathcal S_\mu
\equiv
-u_\mu
-i(S\cdot k)_\mu \,\widehat{\mathscr D}_\xi\,.
\end{equation}
We will define $\widehat{\mathscr D}_\xi$ momentarily, but to match \eqref{eq:MampA} and \eqref{eq:NampA} we require only that it acts on functions of $\xi$ as
\begin{equation}
\label{eq:Dxif}
\widehat{\mathscr D}_\xi\, f(\xi)
=
\mathscr D_\xi \,f(\xi)\,, \qquad 
\mathscr D_\xi=\frac1{\xi}\frac{d}{d\xi}\,.
\end{equation}
In order to interpret $\mathcal S_\mu$ as a spin-raising operator, its defining property must be
\begin{equation}
\label{eq:HSamp}
\mathcal A_{\text{spin-}s}^X(\varepsilon,k,u)
=
\varepsilon^{\mu_1\cdots\mu_s}\,
\mathcal S_{\mu_1}\cdots
\mathcal S_{\mu_s}\,
\mathcal A^X_{\text{spin-}0}(k,u)\,,
\end{equation}
where we again employ $X$ to mean that the expression applies to both the $M$ and $N$ sectors.
We have two clues, namely the case $s=2$ for $D\geq4$, and the case $D=4$ for any $s$. Focusing now on the former clue, the conditions for \eqref{eq:HSamp} to reproduce the graviton amplitudes \eqref{eq:Mamph} and \eqref{eq:Namph} are
\begin{equation}
\left[\widehat{\mathscr D}_\xi\,,\, u_\mu\right]=0\,,
\qquad
\left[(S\cdot k)_{(\mu}
\widehat{\mathscr D}_\xi\,,\,(S\cdot k)_{\nu)}
\right]
=
u_\mu u_\nu\,.
\end{equation}
These conditions should be seen as a commutation rule, for which we currently do not have a natural geometric interpretation.
Together with \eqref{eq:Dxif}, they define the action of the operator $\widehat{\mathscr D}_\xi$, and therefore define the spin-raising operator $\mathcal S_{\mu}$. Based on this definition, we guess the following closed-form combinatorial expression for $\mathcal A_{\text{spin-}s}^X$. Using the shorthand notation
\begin{equation}
\varepsilon\,u^q(Sk)^r
=
\varepsilon_{\mu_1\cdots\mu_{q+r}}\,
u^{\mu_1}\cdots u^{\mu_q}\,
(S\cdot k)^{\mu_{q+1}}\cdots
(S\cdot k)^{\mu_{q+r}}\,,
\end{equation}
we have
\begin{align}
\mathcal A_{\text{spin-}s}^X
=
\sum_{r=0}^{s}
\sum_{p=0}^{\lfloor r/2\rfloor}
&
\frac{s!}{(s-r)!\,(r-2p)!\,2^p p!}
(-1)^{s-r}
(-i)^r
\nonumber\\
&
\times
\varepsilon\,u^{s-r+2p}\,(Sk)^{r-2p}\,
\mathscr D_\xi^{\,r-p}
\mathcal A^X_{\text{spin-}0}\,.
\end{align}
For $s=2$, this formula reproduces the graviton amplitudes \eqref{eq:Mamph} and \eqref{eq:Namph}. For $s=3$, it gives
\begin{align}
\mathcal A^{X}_{\text{spin-}3}
=
\Big[
&
\varepsilon_{\mu\nu\rho}u^\mu u^\nu u^\rho
\left(
-1+3\mathscr D_\xi
\right)
-3i\,\varepsilon_{\mu\nu\rho}
u^\mu u^\nu (S\cdot k)^\rho
\left(
\mathscr D_\xi-\mathscr D_\xi^2
\right)
\nonumber\\
&
+3\,\varepsilon_{\mu\nu\rho}
u^\mu (S\cdot k)^\nu (S\cdot k)^\rho
\mathscr D_\xi^2
+i\,\varepsilon_{\mu\nu\rho}
(S\cdot k)^\mu (S\cdot k)^\nu (S\cdot k)^\rho
\mathscr D_\xi^3
\Big]\,
\mathcal A^X_{\text{spin-}0}\,.
\end{align}
For $s=4$, it gives
\begin{align}
\mathcal A^{X}_{\text{spin-}4}
=
\Big[
&
\varepsilon_{\mu\nu\rho\sigma}
u^\mu u^\nu u^\rho u^\sigma
\left(
1-6\mathscr D_\xi+3\mathscr D_\xi^2
\right)
+4i\,
\varepsilon_{\mu\nu\rho\sigma}
u^\mu u^\nu u^\rho (S\cdot k)^\sigma
\left(
\mathscr D_\xi-3\mathscr D_\xi^2
\right)
\nonumber\\
&
-6\,
\varepsilon_{\mu\nu\rho\sigma}
u^\mu u^\nu (S\cdot k)^\rho (S\cdot k)^\sigma
\left(
\mathscr D_\xi^2-\mathscr D_\xi^3
\right)
-4i\,
\varepsilon_{\mu\nu\rho\sigma}
u^\mu (S\cdot k)^\nu (S\cdot k)^\rho (S\cdot k)^\sigma
\mathscr D_\xi^3
\nonumber\\
&
+
\varepsilon_{\mu\nu\rho\sigma}
(S\cdot k)^\mu(S\cdot k)^\nu(S\cdot k)^\rho(S\cdot k)^\sigma
\mathscr D_\xi^4
\Big]\,
\mathcal A^X_{\text{spin-}0}\,.
\end{align}

We can now test this proposal with our other clue, namely the spin-$s$ case in $D=4$. This builds on the discussion in section~\ref{sec:4Damps}. Using the helicity polarisation tensors
\begin{equation}
\varepsilon^\pm_{\mu_1\cdots\mu_s}
=
\epsilon^\pm_{\mu_1}\cdots\epsilon^\pm_{\mu_s}\,,
\end{equation}
we find that the explicit examples up to spin 6 are consistent in $D=4$ with the formula for $s\geq1$
\begin{align}
\mathcal A^M_{\text{spin-}s}(\varepsilon^\pm)
& =
(-1)^s
(\epsilon^\pm\cdot u)^s
\left(
1\mp i\frac{d}{d\xi}
\right)^s
\mathcal A^M_{\text{spin-}0}
\nonumber \\
& = (-1)^s\,2^{s-1}\,8\pi^2\,
(\epsilon^\pm\cdot u)^s\,
e^{\mp a\cdot k}
\,.
\end{align}
The $N$-sector amplitude follows similarly, leading to
\be
\mathcal A^N_{\text{spin-}s}(\varepsilon^\pm)
=
\pm\mathcal A^M_{\text{spin-}s}(\varepsilon^\pm)\,.
\ee
Hence,
\be
\label{eq:HSMNamp}
M\,\mathcal A^M_{\text{spin-}s}(\varepsilon^\pm)
+
N\,\mathcal A^N_{\text{spin-}s}(\varepsilon^\pm)
=
(-1)^s\,2^{s-1}\,8\pi^2\,
(\epsilon^\pm\cdot u)^s
(M\pm N)
e^{\mp a\cdot k}\,.
\ee
Our proposed higher-spin amplitudes match the expectations for helicity amplitudes in $D=4$ \cite{Arkani-Hamed:2017jhn,Guevara:2018wpp,Huang:2019cja}.

The spin-raising operator \eqref{eq:spinraising} that defines the higher-spin amplitudes is one of the main results of this paper.


\section{From amplitudes back to solutions}
\label{sec:tests}

In this section, we will translate the structure of the scattering amplitudes from momentum space to position space, using the map from the amplitudes to the linearised curvatures. This will allow us to write expressions for the spin-$s$ Kerr-NUT curvatures in terms of the position-space scalar seeds $\Phi_M$ and $\Phi_N$, mirroring the story for the amplitudes. Such expressions facilitate the comparison to the original position-space construction from section~\ref{sec:KTN}, which was based on the multi-Kerr-Schild gauge. Further simplifications occur when we take the equal-rotations limit, and focus on the simpler `little duality' for even $D$, building on section \ref{sec:KTNeqspin}.

\subsection{Kerr-NUT fields from potentials}
\label{sec:potentials}

Our first step is to express the linearised curvatures obtained from the amplitudes via \eqref{eq:hsRA} in a simpler manner, which will facilitate the comparison. For this purpose, we introduce dual towers of potentials: using $\sigma=\frac{D-5}{2}$,
\begin{align}
\Theta^{(s)}_{M} &\equiv (-\mathscr D_\chi)^s\,\Phi_M  =
2^\sigma\,\Gamma(\sigma+1)\,
\frac{J_{\sigma+s}(\chi)}{\chi^{\sigma+s}}\,
\frac1{R^{D-3}}\,, \\
\Theta^{(s)}_{N} &\equiv (-\mathscr D_\chi)^s\,\Phi_N=
(-1)^s\,\eta_D\,2^\sigma\,\Gamma(\sigma+1)\,
\frac{J_{-(\sigma+s)}(\chi)}{\chi^{\sigma+s}}\,
\frac1{R^{D-3}}\,.
\end{align}
As in \eqref{eq:A0N}, we define $\eta_D$ to be $i(-1)^{\frac{D}{2}}$ in even $D$ and $(-1)^{\frac{D-3}{2}}$ in odd $D$. The towers start at $s=0$, with the scalars $\Phi_M$ and $\Phi_N$ we first encountered in section~\ref{sec:KTN}, and the higher-$s$ elements are obtained by acting with the position-space Bessel-order shifting operator
\be
\mathscr D_\chi \equiv \frac1{\chi}\,\frac{d}{d\chi}\,,
\qquad
\chi^{2} = -\sum_{i} a_i^{2}\left(\partial_{x_i}^{2}+\partial_{y_i}^{2}\right)\,.
\ee
We have
\be
\label{eq:oddDTheta}
\text{in odd $D$,} \qquad
\Theta_N^{(s)}+\Theta_M^{(s)}=0\,.
\ee
In even $D$, the two towers are instead the real and imaginary pieces of a single object:
\be
\label{eq:ThetaH}
\text{in even $D$,} \qquad
\Theta_M^{(s)}+\Theta_N^{(s)}
= (-\mathscr D_\chi)^s\,(\Phi_M+\Phi_N) =
2^\sigma\Gamma(\sigma+1)\,
\frac{H_{\sigma+s}^{(1)}(\chi)}
{\chi^{\sigma+s}}
\frac1{R^{D-3}}\,,
\ee
where $H_{\rho}^{(1)}$ denotes a Hankel function. This generalises the $s=0$ case of \eqref{eq:PhiH}. We note that the $\Theta^{(s)}_{M}$ are power series in $\chi^2$ for any $D\geq4$, and hence so are the $\Theta^{(s)}_{N}$ for odd $D$. For even $D$, however, we have $\Theta^{(s)}_{N}\propto\chi^{5-D-2s} R^{-(D-3)}$ in the small-rotations limit. The action of negative powers of $\chi$ should be understood in Fourier space. As Fourier integrals, the potentials are
\be
\label{eq:Thetas}
\Big(\Theta^{(s)}_{M}\,,\, \Theta^{(s)}_{N} \Big)
=
(2\pi)^{\frac{D+1}{2}}
\int\frac{d^Dk}{(2\pi)^D}\,
\frac{e^{ik\cdot x}}{k^2}\,
\frac{\delta(k\cdot u)}{\xi^{\,s+\frac{D-5}{2}}}\,
\Big(
J_{s+\frac{D-5}{2}}(\xi)\,,
\,(-1)^s\,\eta_D\,
J_{-\left(s+\frac{D-5}{2}\right)}(\xi)
\Big)\,.
\ee
These are the elementary Fourier integrals needed to express the right-hand side of the curvature/amplitude map \eqref{eq:hsRA} for our spin-$s$ amplitudes. The issue of whether these integrals converge will be discussed at the end of this section.

It is helpful to adopt the rest frame, where the velocity is $u^\mu=\delta^\mu_t$ and the non-vanishing components of the (rescaled) angular momentum are $S_{x_iy_i}=-S_{y_ix_i}=a_i$, for $i=1,\cdots, m$\,; recall \eqref{eq:Suv}. We will also label the coordinates so that the $x^I$ are the spatial ones:
\be
(t,x_1,y_1,\cdots,x_m,y_m,z) = (x^0,x^I)\,,
\ee
with the coordinate $z$ being absent in odd $D$.

Starting with spin 0, we have simply
\be
\Phi_X=\Theta^{(0)}_{X}\,, \qquad \text{where} \quad X={M,N}\,.
\ee
For spin 1, we have $F_{\mu\nu}^X= \mathcal R^{\text{spin-}1,X}_{\mu\nu}$. By pulling factors $k_I$ out of a Fourier integral as $-i\partial_I$, the expressions \eqref{eq:FA}, \eqref{eq:MampA} and \eqref{eq:NampA} lead to
\be
\mathcal R^{\text{spin-}1,X}_{0I} = - \partial_I \Phi_X = - \partial_I \Theta^{(0)}_{X}\,,
\qquad
\mathcal R^{\text{spin-}1,X}_{IJ} = 2\, (S\cdot\partial)_{[J} \partial_{I]} \Theta^{(1)}_{X}\,,
\ee
where we use the short-hand notation $(S\cdot\partial)_I = S_I{}^J\partial_J$. Notice that the equation for $\mathcal R^{\text{spin-}1,X}_{0I}$ is straightforwardly verified from the multi-Kerr-Schild form of the field, because the time component of $\ell_{(\alpha)}$ is 1 for any $\alpha$. An analogous statement holds for curvatures of any spin $s$ with at least one time component:
\be
\mathcal R^{\text{spin-}s,X}_{0I|\mu_2\nu_2|\cdots|\mu_s\nu_s}
=
-\partial_I
\mathcal R^{\text{spin-}(s-1),X}_{\mu_2\nu_2|\cdots|\mu_s\nu_s}\,.
\ee
For this reason, to verify the curvature/amplitude relation at spin $s$, it is sufficient to consider only spatial components of the curvatures, assuming that the curvatures for spin $<s$ have already been checked. Moving on to spin 2, the purely-spatial components of the curvature as computed from the curvature/amplitude relation are
\be
\mathcal R^{\text{spin-}2,X}_{I_1J_1|I_2J_2}
=
4\left[
(S\cdot\partial)_{J_1}(S\cdot\partial)_{J_2}
\partial_{I_1}\partial_{I_2}\Theta_X^{(2)}
+
\delta_{J_1J_2}
\partial_{I_1}\partial_{I_2}\Theta_X^{(1)}
\right]_{[I_1J_1],[I_2J_2]}\,,
\ee
where the final subscript indicates antisymmetrisations performed separately in each
pair $(I_a,J_a)$, to avoid the clutter of the usual brackets.
One can proceed to higher spins, arriving at the expression
\be
\mathcal R^{\text{spin-}s,X}_{I_1J_1|\cdots|I_sJ_s}
=
2^s\!\left[
\sum_{q=0}^{\lfloor s/2\rfloor}
\sum_{\mathcal P_q}
\left(
\prod_{\{a,b\}\in\mathcal P_q}
\delta_{J_aJ_b}
\right)\!
\left(
\prod_{c\notin U(\mathcal P_q)}
(S\cdot\partial)_{J_c}
\right)
\partial_{I_1}\!\cdots\partial_{I_s}
\Theta_X^{(s-q)}
\right]_{[I_1J_1],\cdots,[I_sJ_s]}\,.
\label{eq:RsX}
\ee
Here, $\mathcal P_q$ denotes a set of $q$ disjoint unordered pairs chosen from $\{1,\ldots,s\}$, and $U(\mathcal P_q)$ denotes the set
of labels that appear in those pairs. The second product runs over the unpaired labels. The empty product is understood to be $1$.

We performed numerical checks that these expressions match the curvatures computed from the multi-Kerr-Schild fields up to spin 4 in $D\leq8$, in both the $M$ and the $N$ sectors. A much simpler check is that
\be
\text{in odd $D$,} \qquad
\Theta_N^{(s)}+\Theta_M^{(s)}=0\,, \qquad \text{hence}\qquad 
\mathcal R^{\text{spin-}s,N}_{I_1J_1|\cdots|I_sJ_s} = - \mathcal R^{\text{spin-}s,M}_{I_1J_1|\cdots|I_sJ_s}\,,
\ee
which corresponds to the trivial duality.

Finally, we note that from the expression \eqref{eq:RsX} for the linearised curvatures, and from the definition of the potentials $\Theta_X^{(s)}$, we can read off representatives of the spin-$s$ fields:
\be
\label{eq:psiXpot}
\psi^{X}_{\underbrace{0\cdots0}_{s-r}I_1\cdots I_r}
=
\sum_{q=0}^{\lfloor r/2\rfloor}
\sum_{\mathcal P_q}
\left(
\prod_{\{a,b\}\in\mathcal P_q}
\delta_{I_a I_b}
\right)
\left(
\prod_{c\notin U(\mathcal P_q)}
(S\cdot\partial)_{I_c}
\right)
\Theta_X^{(r-q)} \; + \; \text{pure gauge}.
\ee
Here, $\mathcal P_q$ denotes pairings chosen from the $r$ spatial labels. To conclude, we expressed the spin-$s$ fields in a gauge where they are determined by a spin-raising operation on the scalar seeds $\Phi_M$ and $\Phi_N$ (up to the caveat in the following paragraph), mirroring the amplitudes story. Moreover, we note that for odd $D$ the expressions for the spin-$s$ fields in the new gauge explicitly obey the trivial duality.

We have smoothed over an important point in this section, namely whether the Fourier integrals \eqref{eq:Thetas} converge, due to the behaviour at small $\xi$. The dangerous cases occur for the $N$-sector potentials $\Theta_N^{(s)}$ in even $D$. However, we have shown that the associated $N$-sector curvatures \eqref{eq:RsX} have convergent integrals, at least for $z\neq0$, as described in appendix~\ref{app:Nconv}. So the gauge-invariant information is well defined, and we can verify the matching between the curvatures computed from the potentials and the curvatures computed from the multi-Kerr-Schild fields, which is precisely the check we have made. On the other hand, the Fourier integrals for the potentials $\Theta_N^{(s)}$ and for the representatives \eqref{eq:psiXpot} of the $N$-sector spin-$s$ fields $\psi^N_{\mu_1\cdots\mu_s}$ are not generally convergent, and therefore require a finite-part prescription. An analysis similar to that in the appendix indicates only that the potentials and the representatives of the fields (meaning all their components) are guaranteed to converge for $s=0$ and $s\leq1$, respectively.


\subsection{Tests for equal rotations in even $D$}
\label{sec:eveneqa}

Now we focus on semi-analytic checks for even $D$ in the equal-rotations case, $a_i^2=a^2$, which we first explored in section~\ref{sec:KTNeqspin}. We discussed there that, beyond the general duality relating the $M$ and $N$ sectors, for equal rotations there is a `little duality' relating only the sectors obtained from the solutions $r_{(0)}$ and $r_{(1)}$ of the spheroidal radial equation. The difference between the two settings lies in the zero-modes of the operators $\partial_z^n$. Dropping these zero modes, it is much easier to proceed analytically up to the stage where we are comparing local expressions, with no integrals left. The final numerical comparison between local expressions is straightforward.

In the rest frame, with $k=(k_0,\vec{k}_\perp,k_z)$, we have for equal rotations $\xi=|a\,\vec{k}_\perp|$. However, up to the zero-modes of $\partial_z^n$, we can take $\xi$ to be $iak_z$, which is the same choice of branch as taking $\chi$ to be $-a\partial_z$ in section~\ref{sec:KTNeqspin}. This is the step where the full $N$ sector is truncated to the $r_{(1)}$ sector, so that the setting is now the `little duality'. The relevant potentials are then
\be
\Theta_{(0)}^{(s)}\equiv(-\mathscr D_{-a\partial_z})^s \,\Phi_{(0)}=\Theta_{M}^{(s)}\,, \qquad
\Theta_{(1)}^{(s)}\equiv(-\mathscr D_{-a\partial_z})^s \,\Phi_{(1)}\,,
\ee
where $\Theta_{(1)}^{(s)}\neq\Theta_{N}^{(s)}$ for $D\geq6$.
Recalling that the $\Theta_{(0)}^{(s)}$ are real and the $\Theta_{(1)}^{(s)}$ are imaginary in even $D$, it is useful to package them into a single tower of complex potentials:
\be
\label{eq:XsHankel}
X_s
\equiv
\Theta_{(0)}^{(s)}+\Theta_{(1)}^{(s)} =
(-\mathscr D_{-a\partial_z})^s
\big(\Phi_{(0)}+\Phi_{(1)}\big)=
2^\sigma\Gamma(\sigma+1)\,
\frac{H^{(1)}_{\sigma+s}(-a\partial_z)}
{(-a\partial_z)^{\sigma+s}}\,
\frac1{R^{D-3}}\,.
\ee
This is the higher-spin version of the Newman-Janis-like
formula \eqref{eq:Phi01}:
\be
X_s
=
e^{-ia\partial_z}\,
f^{(D)}_s(-a\partial_z)\,
\frac1{R^{D-3}}\,,
\ee
where \(f_s^{(D)}\) is a polynomial in inverse powers of its argument. Inverse powers of \(\partial_z\) require a zero-mode prescription, as discussed in section~\ref{sec:KTNeqspin}.

Let us now make this explicit in $D=6$. We define
\be
\varrho=|\vec x_\perp|\,,
\qquad
w=z-ia\,,
\qquad
R_w=\sqrt{\varrho^2+w^2}\,,
\qquad
\Upsilon=\operatorname{arcsinh}\!\left(\frac{w}{\varrho}\right).
\ee
We also introduce a tower of $z$-antiderivatives of $R_w$, that is, of the Newman-Janis-shifted $R$:
\be
G_0=R_w\,,
\qquad
\partial_z G_n=G_{n-1}\,.
\ee
For example,
\be
G_1=
\frac{\varrho^2}{2}\Upsilon+\frac{w}{2}R_w\,,
\qquad
G_2=
\frac{\varrho^2w}{2}\Upsilon
+
\left(
\frac{w^2}{6}-\frac{\varrho^2}{3}
\right)R_w\,,
\ee
and
\be
G_3=
\left(
\frac{\varrho^2w^2}{4}-\frac{\varrho^4}{16}
\right)\Upsilon
+
\left(
\frac{w^3}{24}
-\frac{13\varrho^2w}{48}
\right)R_w\,.
\ee
With these definitions, the first few $D=6$ complex potentials are
\be
X_0
=
\frac{i}{a\varrho^2}\frac{w}{R_w}\,,
\qquad
X_1
=
\frac{i}{a^3\varrho^2}
\left(
G_1+iaG_0
\right),
\qquad
X_2
=
\frac{i}{a^5\varrho^2}
\left(
3G_3+3iaG_2-a^2G_1
\right),
\ee
and
\be
X_3
=
\frac{i}{a^7\varrho^2}
\left(
15G_5+15iaG_4-6a^2G_3-ia^3G_2
\right).
\ee
The little-dual potentials are
\be
\label{eq:Theta01}
\Theta_{(0)}^{(s)}
=
\operatorname{Re}X_s\,,
\qquad
\Theta_{(1)}^{(s)}
=
i\operatorname{Im}X_s\,.
\ee
The common curvature expression \eqref{eq:RsX}, evaluated using $\Theta_{(0)}^{(s)}$ and $\Theta_{(1)}^{(s)}$, applies respectively to the $(0)$-sector and $(1)$-sector curvatures. We have explicitly checked in the examples above for $D=6$ up to spin 3 that these curvatures $\mathcal R^{\text{spin-}s,(\alpha=0,1)}_{I_1J_1|\cdots|I_sJ_s}$ 
match the curvatures obtained directly from applying \eqref{eq:hsR} to the Kerr-Schild fields $\psi_{\mu_1\cdots\mu_s}^{(\alpha=0,1)}$.


\section{Self-duality and $2\mapsto 2$ scattering}
\label{sec:SDint}

We unveiled a non-trivial notion of `electric-magnetic' duality of Kerr-NUT fields in even $D$ and for all spins. It relates the electric ($M$) and magnetic ($N$) parts of
\be
\psi_{\mu_1\cdots\mu_s}=M \psi_{\mu_1\cdots\mu_s}^M + N \psi_{\mu_1\cdots\mu_s}^N\,.
\ee
In $D=4$, there is a well-established notion of self-duality, which for these fields corresponds to charges obeying $M=N$; and similarly for anti-self-duality with $M=-N$. We note that this requires a complexification, because our geometric parameters $N$ are $i$ times the physical NUT/magnetic charges in even $D$. Now, as we can verify in \eqref{eq:HSMNamp}, a self-dual/anti-self-dual source only couples to the $+/-$ helicity of the gauge boson in $D=4$. This has a drastic consequence \cite{Doran:2026bng}: two self-dual sources cannot exchange a gauge boson, because the latter must have opposite helicity seen from each source. The result is that self-dual sources do not scatter off each other (at least to leading order, a caveat needed for a gauge boson with self-interactions). This is not unexpected, because the problem can be embedded in a classically-integrable truncation of the interacting theory. In particular, self-dual Yang-Mills theory and self-dual gravity are well-known to be integrable \cite{10.1093/oso/9780198534983.001.0001}. (We note that this instance of integrability is distinct from the also well-known integrability of geodesics on Kerr-NUT metrics for any charges $N_\alpha$ \cite{Frolov:2017kze}; see appendix~\ref{app:KTNLambdaprincipal}. In that case, the probe scattering is not trivial, just `solvable'.)

In this section, we will see that the proxy for classical integrability in $D=4$, namely the triviality of $2\!\mapsto \!2$ scattering, is absent in $D>4$. In particular, given two Kerr-(equal-)NUT `particles' with charges $M_\text{a}$ and $N_\text{a}$, where $\text{a}=1,2$, we will find that there is no non-trivial charge relation that makes the relevant amplitude vanish identically for generic kinematics. The important quantity is the classical $2\mapsto 2$ amplitude,
\begin{equation}
\mathcal A_{2\mapsto 2}
=
\frac{1}{k^2}\, \bar {\mathcal A}_{2\mapsto 2}\,,
\qquad 
\bar {\mathcal A}_{2\mapsto 2} = \sum_{\eta}
\mathcal A_1(\varepsilon^{(\eta)},k,u_1)\,
\mathcal A_2(\varepsilon^{(\eta)},-k,u_2)\,,
\end{equation}
where $k$ is the momentum of the exchanged gauge boson. From this quantity, one can compute the deflection $\Delta u_\text{a}^\mu$ of the particles via a Fourier-type integral, as derived from the KMOC formalism \cite{Kosower:2018adc}. What matters is the residue of $\mathcal A_{2\mapsto 2}$ at $k^2=0$, so $\bar {\mathcal A}_{2\mapsto 2}$ can indeed be factorised as written above, into the product of the 3-point scattering amplitudes we have been studying.

For the sake of simplicity, we will focus here on the gauge theory case, though we will still call the charges $M_\text{a}$ and $N_\text{a}$. The 3-point amplitudes are
\be
\mathcal A_\text{a}(\epsilon^{(\eta)},k,u_\text{a}) = M_\text{a} \,\mathcal A^M(\epsilon^{(\eta)},k,u_\text{a})+ N_\text{a}\, \mathcal A^N(\epsilon^{(\eta)},k,u_\text{a}) = \epsilon^\mu (M_\text{a} \,J^M_{\text{a}\mu}+N_\text{a} \,J^N_{\text{a}\mu})\,.
\ee
Gauge invariance implies current conservation, $k^\mu J^{M,N}_{\text{a}\mu}=0$. The sum over polarisations $\eta$ is evaluated using the completeness relation,
\be
\sum_\eta \epsilon^{(\eta)}_\mu \epsilon^{(\eta)}_\nu = \eta_{\mu\nu} - \frac{k_\mu n_\nu + n_\mu k_\nu}{k\cdot n}\,,
\ee
where $n$ is a null reference vector.
Taking into account current conservation, we have
\be
\label{eq:A4JJ}
\bar {\mathcal A}_{2\mapsto 2} = M_1M_2\, (J^M_1\cdot J^M_2) + M_1N_2 \,(J^M_1\cdot J^N_2) + N_1M_2\, (J^N_1\cdot J^M_2) + N_1N_2 \,(J^N_1\cdot J^N_2)\,,
\ee
where our shorthand $J_1\cdot J_2$ is really $J_1(k)\cdot J_2(-k)$. 
In even $D>4$, there is no linear relation among the four terms, so there is no non-trivial restriction on the charges that eliminates the $2\mapsto 2$ amplitude. However, one can check that
\begin{align}
\label{eq:4DJJ}
\text{in } D=4\,,& \qquad J^M_1\cdot J^M_2=-J^N_1\cdot J^N_2\,, \quad J^M_1\cdot J^N_2 =- J^N_1\cdot J^M_2\,,  \\
 & \qquad \bar {\mathcal A}_{2\mapsto 2} = (M_1M_2-N_1N_2) (J^M_1\cdot J^M_2) + (M_1N_2-N_1M_2) (J^M_1\cdot J^N_2)\,.
 \nonumber
\end{align}
That is, the amplitude depends on the charges only through the two duality-invariant combinations.
Hence, for the self-dual and anti-self-dual cases,
\be
\text{in } D=4\,, \qquad \bar{\mathcal A}_{2\mapsto 2}\big|_{M_\text{a}=N_\text{a}} = 0\,, \qquad \bar{\mathcal A}_{2\mapsto 2}\big|_{M_\text{a}=-N_\text{a}} = 0\,.
\ee
The $J_1\cdot J_2$ identities in \eqref{eq:4DJJ} rely on kinematic relations and on relations among the Bessel functions. Among the kinematic relations, which are valid in $D=4$ under the on-shell constraints, we have
$$
(S_1\cdot k)\cdot (S_2\cdot k)=(u_1\cdot u_2)(k\cdot a_1)(k\cdot a_2)\,, \qquad (k\cdot a_1)\,u_1\cdot(S_2\cdot k)=-(k\cdot a_2)\,u_2\cdot(S_1\cdot k)\,,
$$
with the latter following from the Schouten identity. Regarding the Bessel functions, what distinguishes $D=4$ is the degeneracy $J_{\pm\sigma}=J_{\mp(\sigma+1)}$, with $\sigma=(D-5)/2=-1/2$; this means that only $J_{\pm \frac12}$ appear, leading to straightforward trigonometric identities, whereas in $D>4$ four types of Bessel function appear for the gauge-theory problem. Naturally, in $D=4$, it is much easier to calculate $\bar{\mathcal A}_{2\mapsto 2}$ using the helicity amplitudes, but we wanted to illustrate what is special with respect to the general $D$ structure. As noted above, for $D>4$ the four $J_1\cdot J_2$ terms are linearly
independent as functions of the generic on-shell kinematics, so no charge relation eliminates $\bar{\mathcal A}_{2\mapsto 2}$. We verified this numerically in $D=6$, both for gauge theory and for the analogous problem in gravity. 

We can still define a non-trivial notion of self-duality in higher even dimensions. The duality map $\mathcal A^M \leftrightarrow \mathcal A^N$ means that $M=N$ and  $M=-N$ correspond to self-duality and anti-self-duality, respectively. An equivalent statement is that the self-dual and anti-self-dual parts of the Kerr-(equal-)NUT field are
\be
\psi^{\rm SD}_{\mu_1\cdots\mu_s}=\psi^M_{\mu_1\cdots\mu_s}+\psi^N_{\mu_1\cdots\mu_s}\,,
\qquad
\psi^{\rm ASD}_{\mu_1\cdots\mu_s}=\psi^M_{\mu_1\cdots\mu_s}-\psi^N_{\mu_1\cdots\mu_s}\,.
\ee
As discussed in section~\ref{sec:KTNevenD}, this is distinct from the usual notion of duality arising from the Hodge dual. However, the special features of $D=4$ are not present. In particular, we do not expect the self-duality constraint on charges to be associated to an integrable interacting sector of the theory, such as $D=4$ self-dual gravity. Moreover, as already alluded to, the non-vanishing of $\bar{\mathcal A}_{2\mapsto 2}$ for self-dual charges implies that it does not depend on the charges solely through the ($D=4$)-type duality-invariant combinations appearing in \eqref{eq:4DJJ}. Thus, even at the linearised level relevant for single gauge-boson or
graviton exchange, the higher-dimensional duality should not be regarded as a dynamical symmetry of the theory.


\section{Conclusion}
\label{sec:conclusion}

We have described a higher-dimensional extension of the duality between mass/electric and NUT/magnetic charges for Kerr-NUT metrics and their
linearised spin-$s$ multi-copy analogues, and shown that it
is most naturally expressed in momentum space. The central points are that the electric and magnetic sectors are associated with different solutions of the spheroidal radial equation, while their scattering amplitudes are obtained by a spin-raising operation on a scalar seed that exhibits the duality through Bessel order reflection. We also find important differences with respect to $D=4$, illustrating what survives of the Newman-Janis shift and how the $D=4$ self-dual
integrability diagnostic fails in higher dimensions.

There are many open directions that build on these results. First of all, it could be useful to obtain proofs of various relations for which we have only provided numerical checks --- not just for mathematical rigour, but also because the deeper algebraic understanding that this would require could lead to additional results. For instance, it could help the exploration of the degenerations of the spheroidal radial equation for non-generic rotation parameters, and the associated Kerr-NUT spin-$s$ solutions. We have only touched on this topic with our investigations of the equal-rotations case. One feature is that the equal-NUT/magnetic amplitudes, and therefore the curvatures, generically diverge in higher even dimensions in the limit where the rotation parameters all vanish ($a_i\to0$), if one keeps the charges fixed; an alternative is to rescale the charges to ensure finiteness, and seek an interpretation of the solutions. In addition, we have mostly studied the restricted setting of equal NUT/magnetic charges, which is relevant for the duality, but also found evidence that the case of generic charges is more elaborate.

There are various other open problems. It would be good to have a geometric description of our spin-raising operator $\mathcal S_\mu$, and perhaps to connect our construction to the principal tensor. Past work on the spinorial and twistorial interpretations of the classical double copy may provide guidance \cite{Luna:2018dpt,Godazgar:2020zbv,White:2020sfn,Chacon:2021wbr,Ortaggio:2023cdz,Zhao:2024ljb,Zhao:2024wtn,Easson:2023dbk,Easson:2026xod}. In four-dimensional gravity, the duality between the mass and the NUT charge can be understood as an Ehlers transformation \cite{Geroch:1970nt,Geroch:1972yt,Ernst:1967wx}; there is a large body of literature on this topic, see e.g.~\cite{Astorino:2023elf,Barrientos:2023tqb} for recent work, as well as \cite{Alawadhi:2019urr,Banerjee:2019saj} in the context of the classical double copy. A natural question is whether our higher-dimensional extension can be described by an Ehlers-like transformation. The relation to the usual notion of electric-magnetic duality in higher dimensions, which involves fields of different tensor type  \cite{Deser:1997mz,Hull:2001iu,Bunster:2013oaa}, is also unclear. Moreover, we find no reason to expect that our duality between stationary solutions extends as a dynamical symmetry of the ordinary photon/graviton linearised fields for $D>4$. In fact, our discussion of $2\!\mapsto\!2$ scattering provides a sharp obstruction to such an extension within the class of sources considered here, because the dependence on the charges is not restricted to the $(D=4)$-type duality-invariant charge combinations. Another question is how our statements on the Newman-Janis shift in $D>4$ fit with the literature, e.g.~\cite{Erbin:2014lwa,Erbin:2016lzq,Mirzaiyan:2017adt,Mirzaiyan:2022ejm,Ortaggio:2023rzp}, and what is the complete picture.

Other related directions, focusing now on scattering amplitudes, include a fuller understanding of the space of gravitational 3-point amplitudes in higher dimensions, beyond the known Myers-Perry (and now also dual equal-NUT) cases, building on \cite{Gambino:2024uge,Bianchi:2025xol,Campanella:2026wqt,Pokraka:2024fao,Cangemi:2026cjy,Monteiro:2018xev}. The space of vacuum black-hole solutions is famously much richer in higher dimensions \cite{Emparan:2007wm,Emparan:2008eg,Emparan:2009at,Emparan:2013moa}, with an intricate pattern of additional horizon topologies, e.g.~\cite{Emparan:2001wn,Elvang:2007rd,Chen:2008fa}, in some cases related to instabilities of the Myers-Perry branch in the `ultraspinning' limit \cite{Emparan:2003sy,Dias:2009iu,Dias:2014cia}. It would be interesting to know whether our spin-raising operator plays a role beyond the Kerr-NUT family. Besides its intrinsic interest, higher-dimensional gravity is relevant to four-dimensional physics due to the practical use of dimensional regularisation in perturbative calculations, as explored recently in \cite{Akpinar:2025huz}. Moving beyond 3-point amplitudes, there is now a substantial literature on Compton amplitudes for Kerr black holes (massive line $+$ two gravitons) and their use in post-Minkowskian dynamics
and waveform calculations, e.g.~\cite{Chung:2018kqs,Aoude:2022thd,Bautista:2022wjf,Bautista:2023sdf,Alessio:2023kgf,Cangemi:2023bpe,Bjerrum-Bohr:2023jau,Ben-Shahar:2023djm,Brandhuber:2023hhl,Kim:2024grz}. This is closely connected to broader effective field theory, amplitudes and worldline approaches to spinning compact-body dynamics
\cite{Porto:2005ac,Levi:2015msa,Bern:2020buy,Jakobsen:2021lvp}. Myers-Perry Compton amplitudes remain unexplored. There are also various closely-related works \cite{Gonzo:2021drq,Ball:2023xnr,Gonzo:2024zxo,Alessio:2025flu,Akpinar:2025tct} on the integrability of geodesic motion and beyond, using lessons from amplitudes and the double copy, which should have natural higher-dimensional extensions. Yet another direction of interest is to introduce the cosmological constant $\Lambda$, which is relevant to both cosmology and anti-de-Sitter holography. We have maintained $\Lambda$ in appendix \ref{app:KTNLambda} for future reference if one seeks analogues of the flat-space scattering amplitudes we studied. 

Finally, the NUT charge has recently featured in celestial holography, in particular in the self-dual black hole \cite{Crawley:2021auj}; see e.g.~\cite{Crawley:2023brz,Adamo:2023fbj,Adamo:2025fqt,Guevara:2023wlr,Guevara:2024edh,Kim:2024dxo,Kim:2024mpy,Skvortsov:2025ohi,Kim:2025xka,Adamo:2026obu,Doran:2026bng}. As we illustrated when studying $2\!\mapsto\!2$ scattering, the remarkable properties of self-duality in $D=4$ do not typically extend to higher dimensions. Nevertheless, the notion of self-duality we introduced in even $D>4$ may still prove useful.

\section*{Acknowledgements}
We thank Riccardo Gonzo, Cynthia Keeler, Axel Kleinschmidt, Fabio Riccioni and Chris White for discussions. Some of the numerical cross-checks and convergence analyses in this paper were carried out with the assistance of ChatGPT and Claude. All final results and interpretations are the responsibility of the authors. RM and LR are supported by the Royal Society, via a University Research Fellowship and a Newton International Fellowship, respectively. This work is supported by the UK's Science and Technology Facilities Council (STFC) Consolidated Grants ST/T000686/1 and ST/X00063X/1 ``Amplitudes, Strings \& Duality".

\appendix

\section{Extension to Kerr-NUT-(A)dS}
\label{app:KTNLambda}

Here, we extend various equations presented in section~\ref{sec:KTN} to include the cosmological constant, make some remarks about the solutions to the spheroidal radial equation, and state the relation of our coordinates to those used in the previously known multi-Kerr-Schild form of ref.~\cite{Chen:2007fs}.

\subsection{Equations with $\Lambda$}
\label{app:KTNLambdametric}

The Einstein equations are now
\be
R_{\mu\nu}=-\frac{D-1}{L^2} \,g_{\mu\nu}\,,
\ee
where $L$ is the AdS radius, such that $\Lambda=-(D-1)(D-2)/(2L^{2})$. The dS case is trivially obtained by $L^2\mapsto -L^2$.
The Kerr-NUT-AdS solution admits the multi-Kerr-Schild form seen in section~\ref{sec:KTN} but now with base metric
\be
ds^2_{AdS}
=
-\left(1+\frac{R^2}{L^2}\right)dt^2
+
\sum_{i=1}^{m}(dx_i^2+dy_i^2)
+
(1-\eps)dz^2
-
\frac{R^2dR^2}{L^2+R^2}
\,,
\ee
where the Cartesian radius $R$ was given in \eqref{eq:cartR}. The flat limit $L\to\infty$ is manifestly Minkowskian. In addition, we have now
\be
\Phi
=
\frac{
r^{1+\eps}
}{
\prod_{i=1}^{m}(r^2+a_i^2)
}
\left(
1-
\sum_{i=1}^{m}
\frac{
\Xi_i a_i^2(x_i^2+y_i^2)
}{
(r^2+a_i^2)^2
}
\right)^{-1}\,, \qquad  \Xi_i\equiv1-\frac{a_i^2}{L^2}
\ee
\be
\ell =\frac{L^2+R^2}{L^2+r^2}dt + \frac{L^2}{L^2+r^2} \left[(1-\epsilon)\frac{zdz}{r} 
+ \sum_{i=1}^m \frac{\Xi_ir(x_idx_i+y_idy_i)}{r^2+a_i^2}\right]
+ \sum_{i=1}^m \frac{a_i(x_idy_i-y_idx_i)}{r^2+a_i^2}\,,
\ee
The AdS-deformed spheroidal radial equation is
\be
\sum_{i=1}^{m}
\frac{\Xi_i(x_i^2+y_i^2)}{r^2+a_i^2}
+
(1-\eps)\frac{z^2}{r^2} =1 \,.
\ee
The analogue of \eqref{eq:Phi0J} is
\begin{equation}
\Phi_{(0)} = 2^{\sigma}\,\Gamma(\sigma+1)\,
\frac{J_{\sigma}(\chi)}{\chi^{\sigma}}\,\frac{1}{\tilde R^{D-3}}\,,
\qquad
\sigma=\frac{D-5}{2}\,,
\qquad
\chi^{2} \equiv -\sum_{i} \frac{a_i^{2}}{\Xi_i}\left(\partial_{x_i}^{2}+\partial_{y_i}^{2}\right)\,,
\end{equation}
now with \,$\tilde R^2\equiv\sum_{i=1}^m \Xi_i(x_i^2+y_i^2) \,+ (1-\epsilon)z^2\,.$

The statements on the double-copy structure extend naturally. In particular, the single copy \eqref{eq:A} solves the Maxwell equations on both AdS and Kerr-NUT-AdS, as shown in \cite{Krtous:2007xg} in a different coordinate system. For the scalar \eqref{eq:phi}, we have now $\big(\square_\text{AdS}+2(D-3)/L^2\big)\phi=0$\,. See e.g.~ \cite{Bahjat-Abbas:2017htu,Carrillo-Gonzalez:2017iyj} for a discussion of the classical double copy on (A)dS. Analogous statements for linearised higher-spin fields in AdS, where the Fronsdal equation can be covariantised while maintaining gauge invariance, can be found in \cite{Didenko:2022qxq}.

The statements on the reality conditions and the odd-$D$ parameter redundancy also extend. Notice that, with $\Lambda$, a pure-gauge metric perturbation now obeys $W_{\mu\nu\lambda\rho}^{\text{linearised}}=0$, that is, we must consider the Weyl tensor.

The statements about the scalar duality for generic spins in section~\ref{sec:KTN} extend straightforwardly, subject to replacing $1/R^{D-3}$ by $1/\tilde R^{D-3}$ and to considering the $\Lambda$-corrected operator $\chi$. For the many explicit expressions in the even-$D$ equal-spins section, more care is needed, and we cover this in appendix~\ref{app:KTNLambdasol}. The question of how later sections extend to include $\Lambda$ is beyond the scope of this paper.

\subsection{On the solutions $r^2_{(\alpha)}$ of the spheroidal radial equation}
\label{app:KTNLambdasol}

Let us write the spheroidal radial equation in terms of \(u=r^2\):
\be
\label{eq:Fu}
F(u)=0\,, \qquad \text{where}\quad  F(u)\equiv
\sum_{i=1}^m
\frac{\Xi_i(x_i^2+y_i^2)}{u+a_i^2}
+
(1-\epsilon)\frac{z^2}{u}
-1\,.
\ee
$F(u)$ has poles at $u=-a_i^2$ and, in even dimensions, also at $u=0$. In the usual Kerr-AdS range $\Xi_i>0$, $F(u)$ is strictly decreasing on each interval
between its poles:
\be
F'(u)=
-\sum_{i=1}^m
\frac{\Xi_i(x_i^2+y_i^2)}{(u+a_i^2)^2}
-(1-\epsilon)\frac{z^2}{u^2}<0 \,.
\ee
Let us assume that the spins are generic, and let us order them such that $a_i^2<a_{i+1}^2$. The solutions to $F(u)=0$ are such that each lies in an interval as follows:
\begin{align}
\label{eq:inteven}
& \text{for} \quad \eps=0\,, \qquad (-a_m^2,-a_{m-1}^2)\,, \; \cdots\,,\; (-a_2^2,-a_1^2)\,,\; (-a_1^2,0)\,,\; (0,+\infty)\,, \\
\label{eq:intodd}
& \text{for} \quad \eps=1\,, \qquad (-a_m^2,-a_{m-1}^2)\,, \; \cdots\,,\; (-a_2^2,-a_1^2)\,,\; (-a_1^2,+\infty)\,,
\end{align}
with a total of $m+1-\epsilon$ solutions. We denote as $u=r_{(0)}^2$ the largest solution. For even $D$, $r_{(0)}^2>0$. For odd $D$, $r_{(0)}^2>0$ iff $F(0)>0$, that is,~$\sum_{i=1}^m \Xi_i(x_i^2+y_i^2)/a_i^2 >1 $. Picking the positive square root $r_{(0)}$, this corresponds to the Kerr-AdS monomial in the metric, i.e.~the mass monomial. The negative solutions $u=r_{(\alpha)}^2$ lying in the other intervals correspond to $m-\eps$ NUT monomials. Hence, each square root $r_{(\alpha)}$ is imaginary in the NUT cases, so the metric in multi-Kerr-Schild coordinates is complex in Lorentzian signature.

The solutions $r^2_{(\alpha)}$ obey various identities. Let us work with the polynomial
\begin{align}
\tilde F(u) &\equiv - F(u) \, u^{1-\eps}\prod_{i=1}^m(u+a_i^2) \nonumber \\
&= u^{1-\eps}\prod_{i=1}^m(u+a_i^2)
\;-\; u^{1-\eps}\sum_{i=1}^m \Xi_i(x_i^2+y_i^2)\!\!\prod_{j\neq i}\!(u+a_j^2)
\;-\;(1-\eps)\,z^2\prod_{i=1}^m(u+a_i^2)\,.
\label{eq:F}
\end{align}
The highest degree in $u$, which is $m+1-\eps$, comes from the first term above. Hence,
\begin{equation}
\tilde F(u)\;=\;\prod_{\alpha=0}^{m-\eps}\bigl(u-r_{(\alpha)}^2\bigr)\,.
\label{eq:F-fact}
\end{equation}
Expanding \eqref{eq:F} and \eqref{eq:F-fact} in $u$ and matching the coefficients of $u^k$ for $k=0,\ldots,m-\epsilon$ leads to $m+1-\epsilon$ Vieta-type identities. Equivalent identities can be efficiently obtained by equating \eqref{eq:F} and \eqref{eq:F-fact} at the $m$ special points $u=-a_i^2$ and (for even dimensions $\eps=0$) also at $u=0$, leading respectively to:
\be
x_i^2+y_i^2
=
\frac{
\prod_{\alpha=0}^{m-\eps}(r^2_{(\alpha)}+a_i^2)
}{
\Xi_i\,a_i^{2(1-\eps)}
\prod_{j\neq i}(a_i^2-a_j^2)
}\,; \qquad
\text{for $\eps=0$,} \quad z^2
=
(-1)^m
\frac{
\prod_{\alpha=0}^{m}r^2_{(\alpha)}
}{
\prod_{i=1}^{m}a_i^2
}\,.
\label{eq:idxiyiz}
\ee
Another interesting identity, which follows from these, is obtained by taking $u=-L^2$:
\be
R^2
= L^2\left[-1+
\frac1{
\prod_{i=1}^{m}\Xi_i
}\, 
\prod_{\alpha=0}^{m-\eps}\Big(1+\frac{r^2_{(\alpha)}}{L^2}\Big)\right]=
\sum_{\alpha=0}^{m-\epsilon} r_{(\alpha)}^2
+
\sum_{i=1}^{m} a_i^2
+
\mathcal{O}(L^{-2})
\,,
\label{eq:idR}
\ee
where in the last step we considered the $\Lambda\to0$ limit.

Yet another interesting relation is 
\be
\Phi_{(\alpha)}
=
\frac{r_{(\alpha)}^{1-\epsilon}}
{\widetilde F'(u_\alpha)}
=
\frac{r_{(\alpha)}^{1-\epsilon}}
{\displaystyle\prod_{\beta\neq\alpha}(u_\alpha-u_\beta)}
=
\frac{r_{(\alpha)}^{1-\epsilon}}
{\displaystyle\prod_{\beta\neq\alpha}
(r_{(\alpha)}^2-r_{(\beta)}^2)}\,.
\ee
From this relation, it follows by the partial-fractions identity that
\be
\label{eq:redunSa}
\text{for odd $D$}: \quad \sum_{\alpha=0}^{m-1} \Phi_{(\alpha)}=0\,.
\ee
We can also obtain by direct computation that
\be
\label{eq:redunAa}
\text{for odd $D$}: \quad \sum_{\alpha=0}^{m-1}\Phi_{(\alpha)}\ell_{(\alpha)}
=
d\lambda\,,
\ee
with
\be
\lambda
=
\frac{(-1)^{m+1}}
{L^{2m-2}\prod_i\Xi_i}\,t
+
\sum_{i=1}^m
\frac{(-1)^{m+1}a_i}
{\Xi_i\prod_{j\neq i}(a_i^2-a_j^2)}
\,\phi_i
+
\sum_{\alpha=0}^{m-1}
\frac{L^2}{2}
\int^{u_\alpha}
\frac{s_\alpha\sqrt{|u|}\,du}
{(L^2+u)\prod_i(u+a_i^2)}\,,
\ee
where $s_\alpha$ is such that $r_{(\alpha)}=s_\alpha \sqrt{|u_\alpha|}$, with $|s_\alpha|=1$. This is relevant for the single copy discussion in the main text. We conjecture that there is an analogous identity for the double copy at linearised level, but only checked it numerically for $D=5,7$. This identity involves the linearised Weyl tensor,
\be
\label{eq:redunha}
\text{for odd $D$}: \quad \sum_{\alpha=0}^{m-1} W_{\mu\nu\lambda\rho}^{\text{linearised}}\left[\Phi_{(\alpha)}\ell_{(\alpha)\delta}\ell_{(\alpha)\sigma}\right]=0\,.
\ee
Hence, the analogues of \eqref{eq:redunAa} for higher spin are such that the linearised fields are pure gauge. We conjecture that this extends to further higher spins (beyond spin 2).

Another identity is related to reality conditions discussed in the main text. If $r_{(\alpha)}^2<0$, then $i^{1-\epsilon}\,\Phi_{(\alpha)}$ is real, but $\ell_{(\alpha)}$ complex. We have
\be
\text{if\; $r_{(\alpha)}^2<0$\,,} \qquad
\operatorname{Im}\!\left(
i^{1-\epsilon}\,\Phi_{(\alpha)}\ell_{(\alpha)}
\right)
=d\lambda_{(\alpha)}\, ,
\ee
where, up to a numerical factor and a branch-dependent sign,
\be
\lambda_{(\alpha)}
\propto
\int^{u_\alpha}
\frac{L^2 (\sqrt{|u|})^{\epsilon}\,du }
{(L^2+u)\,\prod_i(u+a_i^2)}\,.
\ee
This means that $
d\operatorname{Im}\!\left(
i^{1-\epsilon}\,\Phi_{(\alpha)}\ell_{(\alpha)}
\right)=0$\,.
In addition, we conjecture that
\be
\text{if\; $r_{(\alpha)}^2<0$\,,} \qquad
W_{\mu\nu\lambda\rho}^{\text{linearised}}\left[\operatorname{Im}\!\left(
i^{1-\epsilon}\,\Phi_{(\alpha)}\ell_{(\alpha)\delta}\ell_{(\alpha)\sigma}
\right)\right]=0\,,
\ee
which we checked for $D=4,5,6$.

An important relation involving the $\ell_{(\alpha)}$ is
\be
\label{eq:ellr}
\ell_{(\alpha)}^\mu \partial_\mu r_{(\beta)} = \delta_{\alpha\beta}\,.
\ee
This can be proven by first obtaining an expression for $dr_{(\beta)}$ from $dF(u_\beta)=0$, using $F(u)$ as defined in \eqref{eq:Fu}, and then contracting $dr_{(\beta)}$ with $\ell_{(\alpha)}$. The identity \eqref{eq:ellr} is a crucial step in showing that the field strength $F_{\mu\nu}$ for the single copy \eqref{eq:A} obeys $\ell_{(\alpha)}^\mu F_{\mu\nu}\propto \ell_{(\alpha)\nu}$, which is mentioned in section~\ref{sec:KTNDC}.

\subsection{Solutions $r^2_{(\alpha)}$ for equal spins}
\label{app:KTNLambdasoleq}

Starting from our discussion on solutions for generic spins, in particular the intervals \eqref{eq:inteven} and \eqref{eq:intodd} where the solutions must lie, it is clear what happens when spins coincide. As we take $a_{i+1}^2\to a_i^2$, the solution in the interval $(-a_{i+1}^2,-a_i^2)$ behaves as $r^2\to -a_i^2$. This is generically a singular limit in our expressions for the metric. In the case of strict equality, $a_{i+1}^2=a_i^2$, the spheroidal radial equation has one fewer solution.

The equal-spins case of the spheroidal radial equation is very simple:
\be
a_i^2=a^2\,, \; \forall i \quad \Rightarrow \quad
F(u)=
\frac{\rho^2}{u+a^2}
+
(1-\epsilon)\frac{z^2}{u}
-1\,, \;\;\; \text{where}\;\;\; \rho^2=\Xi\sum_{i=1}^m (x_i^2+y_i^2)\,.
\ee
There are only $2-\eps$ solutions:
\begin{align}
& \text{for} \quad \eps=0\,, \qquad r_{(0)}^2=u_+\,, \; r_{(1)}^2=u_-\,, \quad u_\pm =\frac12\left(\rho^2+z^2-a^2\pm\sqrt{(\rho^2+z^2-a^2)^2+4a^2z^2}\right) \,, \nonumber \\
& {\phantom{\text{for} \quad \eps=0\,, \qquad}} \text{such that} \quad r_{(0)}^2r_{(1)}^2 = -a^2z^2 \,, \\[4pt]
& \text{for} \quad \eps=1\,, \qquad  r_{(0)}^2 = \rho^2-a^2 = \Xi\, R^2-a^2\,.
\end{align}

In even $D$, we have a natural duality between the solutions, with $r_{(0)}$ associated to mass and $r_{(1)}$ associated to NUT charge. 
The discussion in section~\ref{sec:KTNeqspin} for equal spins in even $D$ extends to include $\Lambda$ as follows:
\be
\Phi_{(0)}+\Phi_{(1)} =
2^\sigma\Gamma(\sigma+1)\,
\frac{H_\sigma^{(1)}(-a\partial_z)}
{(-a\partial_z)^\sigma}
\frac1{\tilde R^{D-3}}\,,
\qquad
\sigma=\frac{D-5}{2}\,,
\ee
now with $\tilde R^2=\Xi\sum_{i=1}^m (x_i^2+y_i^2) \,+ z^2.$

In odd $D$, we have a single solution, $r_{(0)}$, associated to mass. Note that the $m-2$ solutions that are lost in this limit ($r^2_{(\alpha>0)}\to -a^2$) are still such that \eqref{eq:redunSa} and \eqref{eq:redunAa} hold.

\subsection{The principal tensor}
\label{app:KTNLambdaprincipal}

For completeness, we mention also the principal tensor. The Kerr-NUT-(A)dS family carries a non-degenerate closed conformal
Killing-Yano tensor called the principal tensor, which we will denote as $H_{\mu\nu}$, whose existence implies algebraic type~D,
geodesic integrability, and separability of standard
field equations on the background~\cite{Frolov:2017kze,Kubiznak:2006kt,Krtous:2006qy}. It satisfies
\begin{equation}
\nabla_\rho H_{\mu\nu} = g_{\rho\mu}\,\xi_\nu - g_{\rho\nu}\,\xi_\mu\,,
\qquad
\xi_\mu \equiv \frac{1}{D-1}\nabla^\nu H_{\nu\mu}\,,
\label{eq:PT-def}
\end{equation}
where $\xi$ is the primary Killing vector. In our coordinates, $\xi=\partial_t$ and
\be
H=db\,, \qquad b
=
-\frac12 \left[R^2\,dt
+\sum_{i=1}^{m}a_i(x_i\,dy_i-y_i\,dx_i)\right]\,.
\ee
The roots $r_{(\alpha)}^2$ of the spheroidal radial equation are related to the principal tensor in that they are the eigenvalues of $H^\mu{}_\lambda H^\lambda{}_\nu$, and one can also show that
\be
H = -\sum_{\alpha=0}^{m-\eps}r_{(\alpha)}\,dr_{(\alpha)}\wedge\ell_{(\alpha)}\,.
\label{eq:Halpha}
\ee
(This is almost the Darboux form of $H$, where instead of $\ell_{(\alpha)}$ we have 1-forms $\hat\omega_{(\alpha)}$ such that $\hat\omega_{(\alpha)}-\ell_{(\alpha)}\propto dr_{(\alpha)}$.)
Integrability of geodesic motion relies on having a tower of $D-1$ commuting symmetries with associated constants of motion. These are made up of $m+1$ Killing vectors (1 time translation and $m$ azimuthal rotations), together with the $m-\eps$ Killing tensors of rank 2 obtained from the principal tensor,
\begin{equation}
K^{(j)}_{\mu\nu} \;=\; \sum_{k=0}^{j}(-1)^k\,e_{j-k}(\{r_{\alpha}^2\})\,
(H^{2k})_{\mu\nu}\,,
\qquad \text{for} \;\;j=1,\dots,m-\eps,
\end{equation}
where we denote $(H^{2k})^\mu{}_\nu=H^\mu{}_{\rho_1}H^{\rho_1}{}_{\rho_2}\cdots H^{\rho_{2k-1}}{}_{\nu}$. The elementary symmetric polynomials $e_{j-k}(\{r_{\alpha}^2\})$ are
precisely the coefficients of $\tilde F(u)$ from \eqref{eq:F} and \eqref{eq:F-fact}:
\be
\tilde F(u)=\sum_{k=0}^{m+1-\epsilon} (-1)^k\, e_k\, u^{m+1-\epsilon-k}\,, \qquad e_0=1.
\ee
The constants of geodesic motion are $\dot x^\mu K_\mu$ and $\dot x^\mu\dot x^\nu K_{\mu\nu}$ for each Killing vector and Killing tensor, respectively. In fact, there are $m+1-\eps$ Killing tensors if we include the metric, $K^{(0)}_{\mu\nu}=g_{\mu\nu}$, but this is associated to a trivial conserved quantity, $g_{\mu\nu}\dot x^\mu\dot x^\nu $.

\subsection{Relation to the Chen-L\"u coordinates}
\label{app:KTNLambdaCL}

The expression \eqref{eq:Halpha} for the principal tensor suggests that the $r_{(\alpha)}$ form an interesting set of variables, geometrically. In fact, this provides the dictionary to the multi-Kerr-Schild coordinates introduced in \cite{Chen:2007fs}, which we will denote with the superscript CL (Chen-L\"u). They use $n$ such that $D=2n$ in even dimension and $D=2n+1$ in odd dimension, which maps to our notation as $n=m+1-\epsilon$. Their setup in signature $(\lfloor D/2\rfloor,\lfloor (D+1)/2\rfloor)=(m+1-\epsilon,m+1)$ is related to our setup in Lorentzian signature via the coordinate map
\be
x_{\alpha+1}^\text{CL} = ir_{(\alpha)} \,, \qquad \tau^\text{CL}=it\,, \qquad \varphi_i^\text{CL}=-i \phi_i\,.
\ee
The $\phi_i$ are our azimuthal coordinates, such that $x_i/y_i=\cot\phi_i$. The mass, NUT and cosmological constant parameters in \cite{Chen:2007fs} are related to ours as
\be
b_{\alpha+1}^\text{CL}=-\frac12
\begin{cases}
i N_\alpha\,, & \epsilon=0\,, \\[2pt]
N_\alpha\,, & \epsilon=1\,,
\end{cases} 
\qquad g^2_\text{CL}=-\Lambda^\text{CL}=L^{-2}\,.
\ee
With these identifications, and using the identities \eqref{eq:idxiyiz} and \eqref{eq:idR}, it is straightforward to identify the Chen-L\"u null one forms as $k_{\alpha+1}^\text{CL} = i\ell_{(\alpha)}$, and to match their metric, namely equation (35) of their paper, to ours. This map makes it clear that the choice of signs for the $r_{(\alpha)}$, which we mentioned at the start, does not affect the multi-Kerr-Schild properties or the Einstein equations, because in the Chen-L\"u coordinates that choice is merely a coordinate choice, where we can take $x_{\alpha+1}^\text{CL}\mapsto -x_{\alpha+1}^\text{CL}$.

We note again that the Kerr-NUT-(A)dS family in $D$ dimensions was introduced by Chen, L\"u and Pope in \cite{Chen:2006xh}, using coordinates that reduce in $D=4$ to the Carter coordinates \cite{Carter:1968ks}. The Chen-L\"u coordinates were then introduced in \cite{Chen:2007fs} as a multi-Kerr-Schild form of the Kerr-NUT-(A)dS metric, which requires a complexification (or change of signature).

\section{Spin-$s$ polarisation sums}
\label{app:projector}

In this appendix, we describe the spin-$s$ polarisation sum
\be
\label{eq:polsum}
P^{(s)}_{\mu_1\cdots\mu_s;\,\nu_1\cdots\nu_s}
\equiv \sum_\eta \varepsilon^{(\eta)}_{\mu_1\cdots\mu_s}\,
\varepsilon^{(\eta)}_{\nu_1\cdots\nu_s}
\ee
following the literature, e.g.~\cite{Behrends:1957rup,Costa:2011mg}. For a complex helicity basis, the second polarisation should be understood as
the complex-conjugate polarisation; we suppress this distinction in the notation. We recall that spin-$s$
polarisations form a basis of symmetric, transverse and traceless rank-$s$
tensors in the SO$(D-2)$ little-group space.
Given the null momentum $k$, we introduce a null reference vector
$n$ such that $k\cdot n\neq0$, and define the elementary transverse projector 
\be
P_{\mu\nu}
\equiv
\eta_{\mu\nu}
-
\frac{k_\mu n_\nu+n_\mu k_\nu}{k\cdot n}\,.
\ee
It obeys
\be
P_{\mu\nu}k^\nu=0\,,
\qquad
P_{\mu\nu}n^\nu=0\,,
\qquad
P^\mu{}_\mu=D-2\,.
\ee
The dependence
on $n$ is the usual gauge dependence of the polarisation
sum, and drops out of gauge-invariant quantities. 

It is convenient to contract the indices of \eqref{eq:polsum} with auxiliary vectors $u$ and $v$, defining
\be
P^{(s)}(u,v)
\equiv
P^{(s)}_{\mu_1\cdots\mu_s;\nu_1\cdots\nu_s}
u^{\mu_1}\cdots u^{\mu_s}
v^{\nu_1}\cdots v^{\nu_s}\,.
\ee
By invariance under the SO$(D-2)$ little group, $
P^{(s)}(u,v)$ must be a polynomial in the invariants
\be
UV \equiv u^\mu P_{\mu\nu}v^\nu\,,
\qquad
UU \equiv u^\mu P_{\mu\nu}u^\nu\,,
\qquad
VV \equiv v^\mu P_{\mu\nu}v^\nu\,.
\ee
Moreover, it must have homogeneity degree $s$ in $u$ and in $v$. Hence,
\be
P^{(s)}(u,v)
=
\sum_{j=0}^{\lfloor s/2\rfloor}
a_j\,
(UV)^{s-2j}\,
(UU)^j\,
(VV)^j\,.
\ee
We choose the normalisation \(a_0=1\). Tracelessness of the polarisation tensors implies that
\be
P^{\mu\nu}\frac{\partial}{\partial u^\mu}\frac{\partial}{\partial u^\nu}\,P^{(s)}(u,v) =0\,.
\ee
This gives the recursion
\be
a_{j+1}
=
-\frac{(s-2j)(s-2j-1)}
{2(j+1)(2s-2j+D-6)}\,a_j\,.
\ee
For the first few spins, we have
\be
P^{(1)}(u,v)=UV\,,
\qquad
P^{(2)}(u,v)
=
(UV)^2
-\frac1{D-2}(UU)(VV)\,,
\ee
and
\be
P^{(3)}(u,v)
=
(UV)^3
-\frac3D(UV)(UU)(VV)\,.
\ee
Equivalently, in index notation,
\be
P^{(1)}_{\mu_1\;\nu_1}
=
P_{\mu_1\nu_1}\,,
\qquad
P^{(2)}_{\mu_1\mu_2;\nu_1\nu_2}
=
\frac12
\left(
P_{\mu_1\nu_1}P_{\mu_2\nu_2}
+
P_{\mu_1\nu_2}P_{\mu_2\nu_1}
\right)
-
\frac1{D-2}
P_{\mu_1\mu_2}P_{\nu_1\nu_2}\,,
\ee
and
\be
P^{(3)}_{\mu_1\mu_2\mu_3;\nu_1\nu_2\nu_3}
=
\mathrm{Sym}
\left[
P_{\mu_1\nu_1}P_{\mu_2\nu_2}P_{\mu_3\nu_3}
\right]
-
\frac3D\,
\mathrm{Sym}
\left[
P_{\mu_1\nu_1}P_{\mu_2\mu_3}P_{\nu_2\nu_3}
\right]\,,
\ee
where $\mathrm{Sym}$ denotes unit-weight symmetrisation separately over the
$\mu$-indices and over the $\nu$-indices.

The argument above uses only the parity-even invariants $UV$, $UU$ and $VV$. In $D=4$, there is an additional parity-odd invariant, $\epsilon_{\mu\nu\rho\sigma}
u^\mu v^\nu k^\rho n^\sigma/k\cdot n$.
This invariant allows one to split the full transverse projector into
positive- and negative-helicity projectors. However, the polarisation sum
$P^{(s)}_{\mu_1\cdots\mu_s;\nu_1\cdots\nu_s}$ used here sums over both
helicities, and is the parity-even projector onto the full symmetric-traceless-transverse space. Therefore the expressions above remain valid
also in $D=4$.


\section{Convergence of Fourier integrals for $N$-curvatures in even $D$}
\label{app:Nconv}

The individual potentials $\Theta_N^{(s)}$ of \eqref{eq:Thetas} grow increasingly
singular with $s$ at small $\xi$. From the Bessel asymptotics, we have in even $D$
\be
\frac{J_{-(\sigma+s)}(\xi)}{\xi^{\sigma+s}}
\;\sim\;
\frac{2^{\sigma+s}}{\Gamma(1-\sigma-s)}\,\xi^{-2(\sigma+s)}\,,
\qquad \text{as} \;\; \xi\to0\,,\qquad \text{with} \;\;\sigma=\frac{D-5}{2}\,.
\ee
The exponent $2(\sigma+s)=D-5+2s$ grows with $s$. For $s\geq1$, the Fourier integral defining $\Theta_N^{(s)}$ already
has a small-$\xi$ singularity requiring a finite-part prescription. Nevertheless, we show here that the gauge-invariant $N$-sector curvatures \eqref{eq:RsX} are free from this ambiguity, at least for $z\neq0$. For generic rotation parameters ($a_i\neq0$), the corresponding Fourier integrals are convergent for every spin $s$ and every even $D$. In fact, this is the case for every term in \eqref{eq:RsX} separately.

In the rest frame, we have $\,\delta(k\cdot u)=\delta(k_t)\,$, and let us split the spatial momentum as $\vec k = (\vec k_\perp,k_z )$, where $\vec k_\perp$ is ($2m$)-dimensional. Let us also denote $\,\xi_A\equiv(S\cdot k)_A\,$, with $A=1,\dots,2m$, so that
\be
|\vec \xi|^2=\sum_i a_i^2\big(k_{x_i}^2+k_{y_i}^2\big)=\xi^2\,.
\ee
For $a_i\neq0$, we have $d^{2m}k_\perp\propto d^{2m}\xi$, because the Jacobian $\prod_i a_i^2$ is constant. The integrand is analytic except where $\vec \xi=0$, that is, on the $k_z$-axis. So we need to investigate the behaviour of the integral near this locus.

First, we perform the $k_z$ integration. In the momentum-space representation of \eqref{eq:RsX}, the $\,(S\cdot k)_{J_c}\,$ are independent of $k_z$, while the derivatives $\partial_{I_1}\!\cdots\partial_{I_s}$ lead to a factor $k_z^{s_z}$, depending on how many indices equal $z$. Gathering the full $k_z$ dependence, we obtain in the patch $z>0$
\be
\int_{-\infty}^{\infty}\!\frac{dk_z}{2\pi}\,
\frac{k_z^{\,s_z}\,e^{ik_z z}}{|\vec k_\perp|^2+k_z^2}
=\tfrac12\,i^{\,s_z}\,|\vec k_\perp|^{\,s_z-1}\,e^{-|\vec k_\perp|z}\,.
\ee
Because $a_i\neq0$, small $|\vec k_\perp|$ is comparable to $\xi$, so the $k_z$ integral provides a factor $\xi^{\,s_z-1}$. Second, we collect the powers of $\xi$ from a single term of \eqref{eq:RsX}, labelled by the pairing $\mathcal P_q$, in an arbitrary spatial component
$\mathcal R^{\text{spin-}s,N}_{I_1J_1|\cdots|I_sJ_s}$:
\begin{itemize}
\item $\Theta_N^{(s-q)}$ gives $\xi^{-2(\sigma+s-q)}=\xi^{-(2m-3+2(s-q))}$;
\item the $k_z$ integral above gives $\xi^{\,s_z-1}$;
\item the $s-s_z$ transverse derivatives $\partial_{I_a}$ give $\xi^{\,s-s_z}$;
\item the $s-2q$ unpaired factors $(S\cdot\partial)_{J_c}$ give $\xi^{\,s-2q}$, because $(S\cdot k)_z=0$.
\end{itemize}
Putting together these powers, we obtain $\xi^{-(2m-2)}$,
which turns out to be independent of $s$, $q$ and $s_z$. Finally, the transverse measure is $d^{2m}\xi=\xi^{2m-1}\,d\xi\,d\Omega_{2m-1}$, so the integral behaves as
\be
\int d\xi\;\xi^{2m-1}\,\xi^{-(2m-2)}=\int d\xi\;\xi
\ee
in the small $\xi$ region, which converges; in fact, the integrand vanishes at $\xi=0$. Our conclusion holds for every term in \eqref{eq:RsX} for $X=N$, irrespective of the
spin, of the even $D=2m+2$, and of the component of the curvature --- as long as $a_i\neq0$. The subleading terms in the small-$\xi$ expansion of the Bessel function are less singular and have improved convergence.

This conclusion for the $N$-sector curvatures does not extend to the potentials, that is, the derivatives in \eqref{eq:RsX} are essential for the improved small-$\xi$ behaviour. We note that the gauge-invariant curvatures are the important objects to compare to the de Wit-Freedman curvatures of the multi-Kerr-Schild fields, and the fact that they match is precisely the check we made numerically. Moreover, we assumed $z\neq0$ for the $k_z$ integral, both in this appendix and in our numerical checks of the curvature formulas. Considering $z=0$ would require a more involved analysis.

\bibliographystyle{apsrev4-1}
\bibliography{mainbib}
\end{document}